\documentclass[aps,prx,twocolumn,amsmath,amssymb,superscriptaddress,10pt,nobibnotes]{revtex4-2}
\usepackage{preamble}

\begin{document}
	

\title{Speckle-correlation imaging through a kaleidoscopic multimode fiber}
\author{Dorian Bouchet}
\altaffiliation{\href{mailto:dorian.bouchet@univ-grenoble-alpes.fr}{dorian.bouchet@univ-grenoble-alpes.fr}}
\affiliation{Universit\'e Grenoble Alpes, CNRS, LIPhy, 38000 Grenoble, France}
\author{Antonio M.\ Caravaca-Aguirre}
\affiliation{Universit\'e Grenoble Alpes, CNRS, LIPhy, 38000 Grenoble, France}
\author{Guillaume Godefroy} 
\affiliation{Universit\'e Grenoble Alpes, CNRS, LIPhy, 38000 Grenoble, France}
\affiliation{Universit\'e Grenoble Alpes, CEA, Leti, 38000 Grenoble, France}
\author{Philippe Moreau}
\affiliation{Universit\'e Grenoble Alpes, CNRS, LIPhy, 38000 Grenoble, France}
\author{Ir\`ene Wang}
\affiliation{Universit\'e Grenoble Alpes, CNRS, LIPhy, 38000 Grenoble, France}
\author{Emmanuel Bossy}
\affiliation{Universit\'e Grenoble Alpes, CNRS, LIPhy, 38000 Grenoble, France}


\begin{abstract}
Speckle-correlation imaging techniques are widely used for non-invasive imaging through complex scattering media. While light propagation through multimode fibers and scattering media share many analogies, reconstructing images through multimode fibers from speckle correlations remains an unsolved challenge. Here, we exploit a kaleidoscopic memory effect emerging in square-core multimode fibers and demonstrate fluorescence imaging with no prior knowledge on the fiber. Experimentally, our approach simply requires to translate random speckle patterns at the input of a square-core fiber and to measure the resulting fluorescence intensity with a bucket detector. The image of the fluorescent object  is then reconstructed from the autocorrelation of the measured signal by solving an inverse problem. This strategy does not require the knowledge of the fragile deterministic relation between input and output fields, which makes it promising for the development of flexible minimally-invasive endoscopes.
\end{abstract}

\maketitle

The development of optical endoscopes is motivated by a number of biomedical applications such as brain imaging~\cite{frank_next-generation_2019}. Multimode fibers are excellent candidates to minimize the invasiveness of such procedures, as they feature a high density of modes per unit area~\cite{snyder_optical_2012,mahalati_resolution_2013}. However, coherent light propagating inside such fibers typically generates speckle patterns at the fiber output, in a similar way as through complex scattering media~\cite{cao_shaping_2022}. Different methods have emerged to exploit the complex deterministic relation between incident and transmitted fields in multimode fibers, based on either experimental measurements~\cite{bolshtyansky_transmission_1996,cizmar_exploiting_2012,choi_scanner-free_2012,papadopoulos_high-resolution_2013,caravaca-aguirre_single_2017,amitonova_endo-microscopy_2020,matthes_learning_2021} or theoretical modeling~\cite{ploschner_seeing_2015,boonzajer_flaes_robustness_2018}. Nevertheless, these methods require either an optical access to both sides of the fiber or a precise knowledge of the fiber geometry over its entire length, which makes them unsuitable for many applications.

In contrast, statistical approaches based on speckle correlations can be implemented without such prior information. In the last decade, these strategies have been successfully employed to image fluorescent objects through layers of scattering materials~\cite{bertolotti_non-invasive_2012,katz_non-invasive_2014,salhov_depth-resolved_2018,wang_non-invasive_2021,zhu_large_2022}. A key component of these approaches is the existence of a memory effect, which creates statistical correlations between incident and transmitted fields~\cite{feng_correlations_1988,freund_memory_1988,schott_characterization_2015,judkewitz_translation_2015,osnabrugge_generalized_2017,yilmaz_angular_2019}. However, multimode fibers have different statistical properties as compared to scattering materials \cite{amitonova_rotational_2015,xiong_long-range_2019,li_memory_2021}. A rotational memory effect exists in commonly-used circular-core multimode fibers~\cite{amitonova_rotational_2015}, but it must be completed by additional information (using e.g. a fluorescent guidestar) in order to form an image~\cite{li_memory_2021}. Recently, deep neural networks have emerged as promising tools to learn not only deterministic but also statistical relations in multimode fibers~\cite{borhani_learning_2018,rahmani_multimode_2018,fan_deep_2019,resisi_image_2021}, but with a generalizability limited to the specific experimental conditions under which the training dataset was measured. Thus, most practical implementations of fiber-optic endoscopes remain currently based on multicore fibers~\cite{andresen_ultrathin_2016,yeminy_guidestar-free_2021,kuschmierz_ultra-thin_2021}, which are characterized by a much larger footprint as compared to multimode fibers.

Traditionally in optical fibers, the geometry of the core is circular, and the use of square-core fibers is limited to specific applications requiring a top-hat-like intensity profile~\cite{velsink_comparison_2021}. However, it was recently observed that a kaleidoscopic memory effect emerges from the strong symmetry properties of square-core multimode fibers~\cite{caravaca-aguirre_optical_2021}. This effect is a special type of shift-shift correlation~\cite{osnabrugge_generalized_2017}: any pattern translating at the fiber input leads to speckle patterns shifting along four directions at the output. The memory effect thus spans the whole two-dimensional (2D) space in square-core fibers, as opposed to conventional multimode fibers with a circular core geometry for which there exists no true radially-shifting memory effect~\cite{amitonova_rotational_2015,li_memory_2021}. Taking advantage of the kaleidoscopic memory effect in square-core fibers, we present here a fully-statistical method to perform endoscopic imaging from speckle correlations, without relying on transmission matrix measurements or on fluorescent guidestars. 

\begin{figure}[ht]
	\begin{center}
		\includegraphics[width=0.95\linewidth]{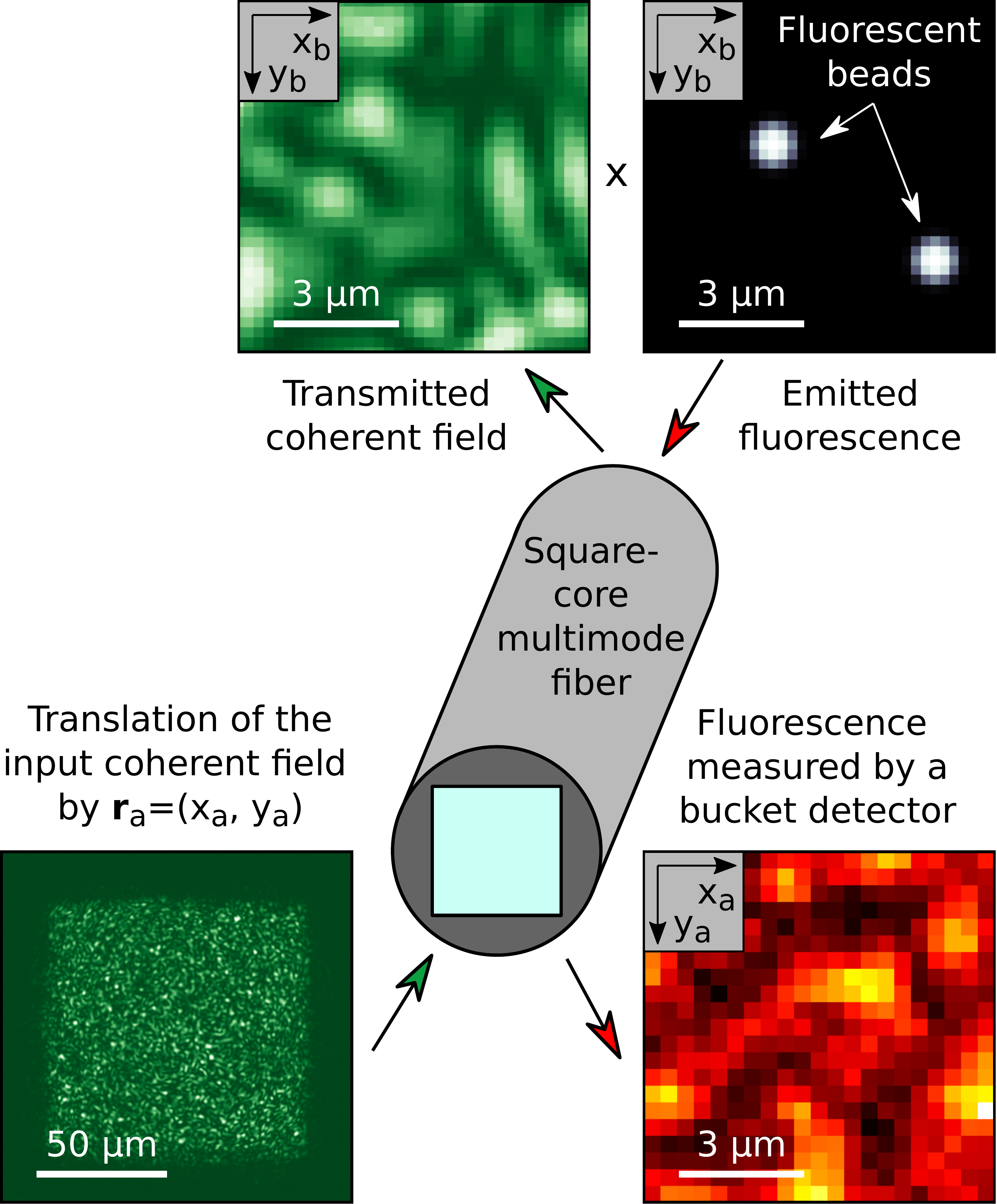}
	\end{center}
	\caption{Principle of a minimally-invasive endoscope based on a square-core multimode fiber. Coherent speckle patterns are translated in the transverse plane at the input of a square-core multimode fiber. Transmitted intensity patterns, which are unknown, excite a fluorescent sample located at the output of the fiber. Fluorescence is collected by the same fiber and is subsequently measured by a bucket detector located at the fiber input. Our method enables one to reconstruct an image of the sample from the seemingly-random fluorescence signal measured as a function of the translation $\bm{r}_{a}$ of the input field.}
	\label{fig_principle}
\end{figure}

\section*{Principle}

The principle of the proposed approach is illustrated in \fig{fig_principle} (see also Methods for a detailed description of the experimental setup). A fluorescent sample is placed at the output of a step-index square-core optical fiber (CeramOptec, core section $100$\,\textmu m\,$\times$\,$100$\,\textmu m). In our proof-of-principle experiment, this sample is composed of several fluorescent beads (ThermoFisher Scientific, red FluoSpheres, diameter $1.0$\,\textmu m). We generate random speckle patterns at the fiber input using a digital micromirror device (DMD), and the transmitted light forms unknown speckle patterns that excite the fluorescent beads. Fluorescence collected through the fiber is then measured by a bucket detector located at the fiber input.

\begin{figure*}[ht]
	\begin{center}
		\includegraphics[width=\linewidth]{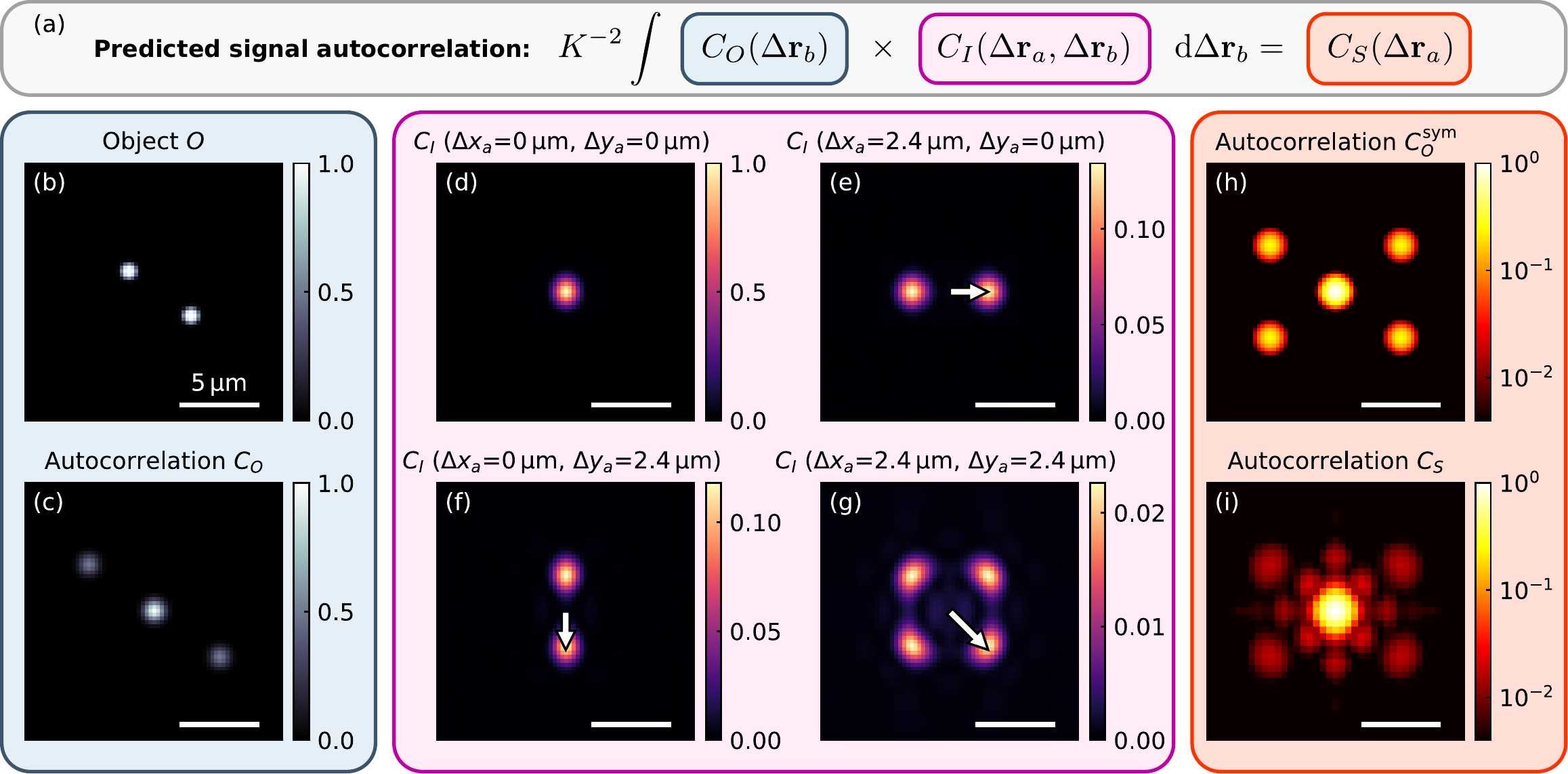}
	\end{center}
	\caption{Predicted autocorrelation of the fluorescence signal. (a) Graphical representation of \eq{eq_corr} relating the object autocorrelation $C_O (\Delta \bm{r}_b)$, the intensity correlation function of the excitation field $C_I (\Delta \bm{r}_a, \Delta \bm{r}_b)$ and the autocorrelation of the fluorescence signal $C_S (\Delta \bm{r}_a)$. (b) Fluorescent object $O(\bm{r}_b)$ composed of two beads (diameter, $1.0$\,\textmu m), that we model here using high-order Gaussian functions. (c) Object autocorrelation $C_O(\Delta \bm{r}_b)$ calculated from $O(\bm{r}_b)$. (d-g) Intensity correlation function of the excitation field $C_I (\Delta \bm{r}_a, \Delta \bm{r}_b)$, represented as a function of $\Delta \bm{r}_b$ (shift at the fiber output) for four different values of $\Delta \bm{r}_a$ (shift at the fiber input). This correlation function, which characterizes light propagation in square-core multimode fibers, is composed of four peaks that translate when changing the value of $\Delta \bm{r}_a$ (white arrows on the figures). (h) Symmetrized autocorrelation of the object $C_O^{\mathrm{sym}}(\Delta \bm{r}_b)$. (i) Predicted signal autocorrelation $C_S(\Delta \bm{r}_a)$ calculated using \eq{eq_corr}. The $4$ main lobes that appear in $C_O^{\mathrm{sym}}(\Delta \bm{r}_b)$ are also identified in $C_S(\Delta \bm{r}_a)$, but with a smaller amplitude due to the finite range of the memory effect. A number of additional lobes can be identified close to the central peak, due to the presence of lobes in the intensity correlation function $C_I (\Delta \bm{r}_a, \Delta \bm{r}_b)$.}
	\label{fig_theory}
\end{figure*}

In order to exploit shift-shift correlations in the fiber, we translate the incident field in the transverse plane and we measure the resulting fluorescence signal $S(\bm{r}_a)$, with a scan area of $8$\,\textmu m\,$\times$\,$8$\,\textmu m. While this signal visually appears as being random, it does carry useful information about the hidden fluorescent object. More precisely, the density distribution of fluorescent emitters $O(\bm{r}_b)$ is related to the measured fluorescence signal $S(\bm{r}_a)$ by the following relation:
\begin{equation}
	S(\bm{r}_a) = \int O(\bm{r}_b) I(\bm{r}_a,\bm{r}_b) \, \de \bm{r}_b \; ,
	\label{eq_signal}
\end{equation}
where $I(\bm{r}_a,\bm{r}_b)$ is the (unknown) excitation intensity transmitted at a position $\bm{r}_b$ at the fiber output for a translation $\bm{r}_a$ of the speckle pattern at the fiber input. From \eq{eq_signal}, we demonstrate in Supplementary Section~2A that the normalized autocorrelation function of the measured fluorescence signal is expressed as follows (see also \fig{fig_theory} for a graphical interpretation of this equation):
\begin{equation}
	C_S (\Delta \bm{r}_a) = K^{-2} \int C_O (\Delta \bm{r}_b) C_I (\Delta \bm{r}_a,\Delta \bm{r}_b) \; \de \Delta \bm{r}_b \; ,
	\label{eq_corr}
\end{equation}
where $C_O (\Delta \bm{r}_b)$ is the autocorrelation of the fluorescent object, $C_I (\Delta \bm{r}_a,\Delta \bm{r}_b)$ is the (known) intensity correlation function of the transmitted excitation intensity, and $K$ is a normalization constant equal to the contrast of the fluorescence signal.

\Eq{eq_corr}, which does not directly involve the object but its autocorrelation (see \fig{fig_theory}b,c), is formally identical to the one describing how to image fluorescent objects through scattering layers~\cite{bertolotti_non-invasive_2012}. However, while the intensity correlation function $C_I$ describing light propagation through scattering layers is characterized by a single peak arising from the usual memory effect~\cite{feng_correlations_1988,freund_memory_1988}, the symmetry of the fiber generates a kaleidoscopic memory effect characterized by a different intensity correlation function~\cite{caravaca-aguirre_optical_2021}. Indeed, $C_I$ is then characterized by four peaks that translate with $\Delta \bm{r}_a$ and that coherently overlap by pair when $\Delta x_a=0$ or $\Delta y_a=0$ (see \fig{fig_theory}d-g and Supplementary Section~3). As a consequence, the autocorrelation of the measured fluorescence signal does not directly yield the object autocorrelation. Instead, it involves $C_O^{\mathrm{sym}} (\Delta \bm{r}_b) = [C_O (\Delta x_b,\Delta y_b)+C_O (\Delta x_b,-\Delta y_b)]/2$, which is a symmetrized version of the object autocorrelation: indeed, since $C_O$ is centrosymmetric, one gets $C_O^{\mathrm{sym}} (-\Delta x_b,\Delta y_b)=C_O^{\mathrm{sym}} (\Delta x_b,-\Delta y_b)$ and $C_O^{\mathrm{sym}} (\Delta x_b,-\Delta y_b)=C_O (\Delta x_b,\Delta y_b)$, as illustrated in \fig{fig_theory}h.

In order to understand how the signal autocorrelation $C_S$ relates to the symmetrized object autocorrelation $C_O^{\mathrm{sym}}$, it is instructive to first consider the limiting case of an infinite-range memory effect with a speckle grain size approaching zero. In this case, \eq{eq_corr} yields $C_S (\Delta \bm{r}_a) \propto w(\Delta \bm{r}_a) C_O^{\mathrm{sym}} (\Delta \bm{r}_a)$, where $w$ is a weight function that is equal to $1$ if $\Delta x_a=0$ and $\Delta y_a=0$, to $1/2$ if either $\Delta x_a = 0$ or $\Delta y_a = 0$, and to $1/4$ otherwise (see Supplementary Section~2B). In practice, the translational memory effect observed in typical step-index square-core optical fibers is further characterized by a limited range~\cite{caravaca-aguirre_optical_2021}, and the measured intensity correlation $C_I (\Delta \bm{r}_a,\Delta \bm{r}_b)$ gradually decays with the distance $\Delta \bm{r}_a$. As a consequence, the signal autocorrelation $C_S (\Delta \bm{r}_a)$ predicted from \eq{eq_corr} also decays with $\Delta \bm{r}_a$ (see \fig{fig_theory}i), limiting the potential reconstruction area to approximately $10$\,\textmu m\,$\times$\,$10$\,\textmu m for the type of step-index fiber used in our experiment. In addition, the finite size of the speckle grain limits the achievable resolution of the method, which is of approximately $1.2$\,\textmu m in our experiment (excitation wavelength $\lambda=532$\,nm, fiber numerical aperture $\mathrm{NA}=0.22$). 

Reconstructing fluorescent objects from measurements of the signal autocorrelation $C_S$ thus amounts to retrieve $O$ from $C_O^{\mathrm{sym}}$. This inverse problem has no known explicit solution, but it can be solved iteratively by exploiting additional prior knowledge about the object as constraints. This problem is in fact deeply connected to commonly-encountered phase-retrieval problems~\cite{fienup_reconstruction_1978,shechtman_phase_2015}, and especially to autocorrelation inversions~\cite{bertolotti_non-invasive_2012}. It essentially differs from such problems in two aspects. First, as the intensity correlation function $C_I (\Delta \bm{r}_a,\Delta \bm{r}_b)$ is characterized by four peaks instead of one, there exists an additional ambiguity in the inverse problem to be solved. Indeed, for a given fluorescent object, there exists four equivalent solutions that are flipped versions of the object (see Supplementary Section~4), instead of two in typical autocorrelation inversions. Second, the signal autocorrelation is weighted by a factor of $1/4$ (instead of $1$) when $\Delta x_a \neq 0$ and $\Delta y_a \neq 0$. This effect, when cumulated to the continuous decay of the correlation function with $\Delta \bm{r}_a$, increases the influence of the statistical fluctuations that are observed when estimating correlation functions from experimental measurements. In practice, to counteract this effect, we generate and translate random incident fields at kHz framerate using the DMD; this allows us to average the signal correlation function over many random configurations of the input speckle, thereby minimizing the apparition of artifacts due to statistical fluctuations. 

\begin{figure}[ht]
	\begin{center}
		\includegraphics[width=\linewidth]{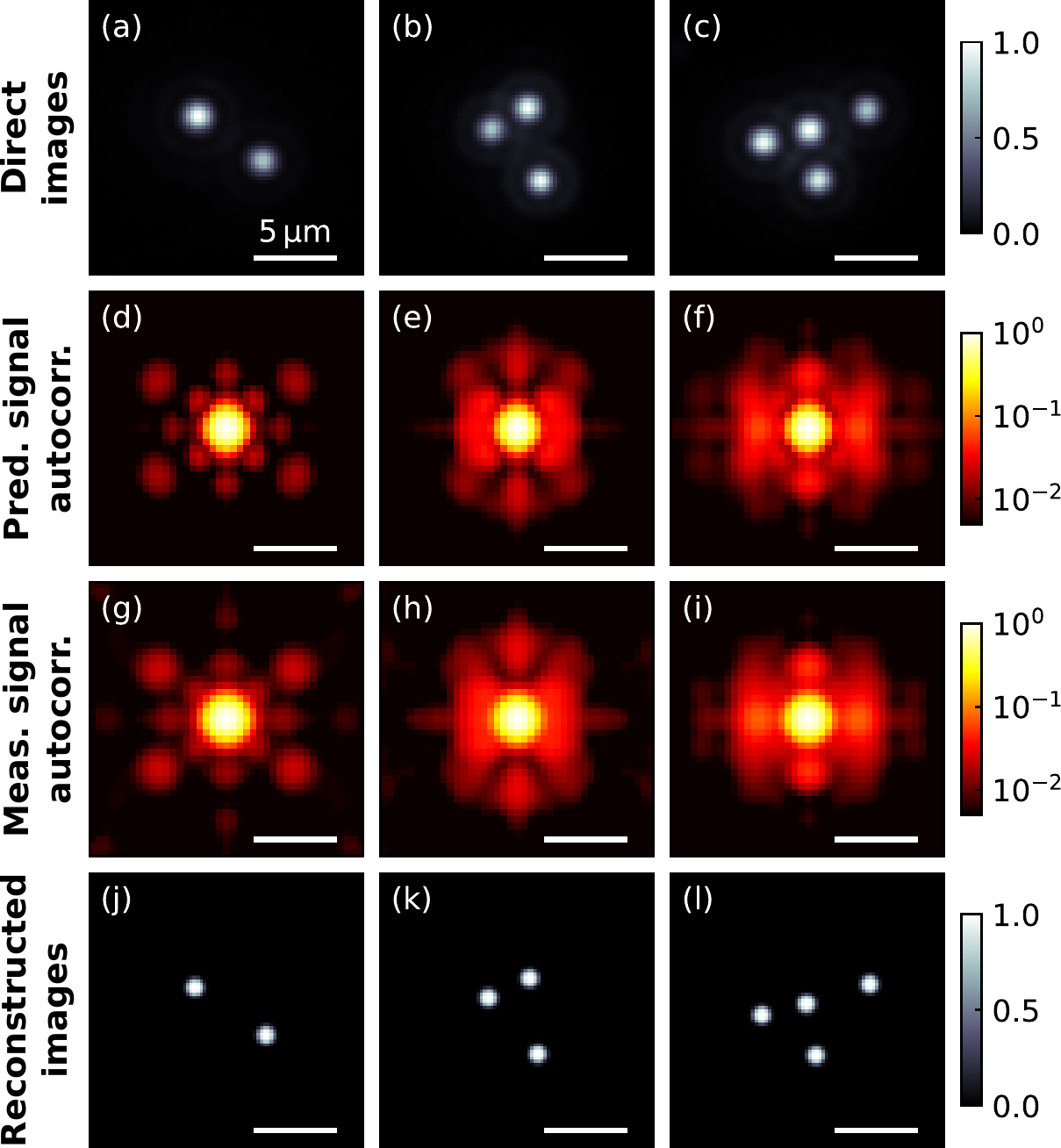}
	\end{center}
	\caption{Reconstructed images of fluorescent beads through a square-core fiber. (a-c) Direct fluorescence images of objects composed of several fluorescent beads, that are located at the output of a $3$\,cm long fiber. (d-f) Predicted signal autocorrelations calculated with the true position of the beads using \eq{eq_corr}. (g-i) Signal autocorrelations retrieved from experimental measurements (averaged over $N_{\mathrm{rep}}=40,000$ speckle illuminations). (j-l) Reconstructed images of the beads. The number of beads is first estimated independently from the statistics of the measured signal (see Supplementary Section~6), and the position of the beads is then retrieved using an optimization algorithm based on simulated annealing.}
	\label{fig_results}
\end{figure}

\section*{Experimental results}

\subsection*{Imaging through a static fiber}

We first experimentally illustrate this approach with a static fiber (fiber length $3$\,cm), that we used to successively probe three different fluorescent objects (\fig{fig_results}a-c). The predicted signal autocorrelation function (\fig{fig_results}d-f), calculated using \eq{eq_corr}, strongly depends on the object, which explicitly demonstrates that this function does carry spatial information about the object. Moreover, autocorrelation functions retrieved from experimental measurements (\fig{fig_results}g-i) are in excellent agreement with theoretically-predicted ones. These results were obtained by averaging over $N_{\mathrm{rep}}=40,000$ realizations of input speckles; while reducing $N_{\mathrm{rep}}$ inevitably leads to the apparition of artifacts due to statistical fluctuations, some spatial information about the fluorescent objects is still present even for much lower values of $N_{\mathrm{rep}}$ (see Supplementary Section~5). Based on these experimental data and on the forward model expressed by \eq{eq_corr}, we could in principle reconstruct an image using a pixel-by-pixel approach. However, the inverse problem is more easily solved when using additional \textit{a priori} information about the object. Here, we take advantage of known characteristics of the beads, setting a bead diameter of $1$\,\textmu m and assuming that all beads have the same brightness. We then estimate the number of fluorescent beads using the statistics of the measured signal (see Supplementary Section~6). Finally, we used an optimization algorithm based on simulated annealing~\cite{kirkpatrick_optimization_1983} in order to find the beads positions that minimize the error between theoretical predictions and experimental data (see Supplementary Section~7). The reconstructed images (\fig{fig_results}j-l) are in excellent agreement with direct images of the objects (i.e. the ground truths), demonstrating that, even though light is apparently scrambled when propagating through square-core multimode fibers, 2D spatial information about fluorescent objects can be effectively recovered with a fully-statistical imaging strategy. 

\begin{figure*}[ht]
	\begin{center}
		\includegraphics[width=\linewidth]{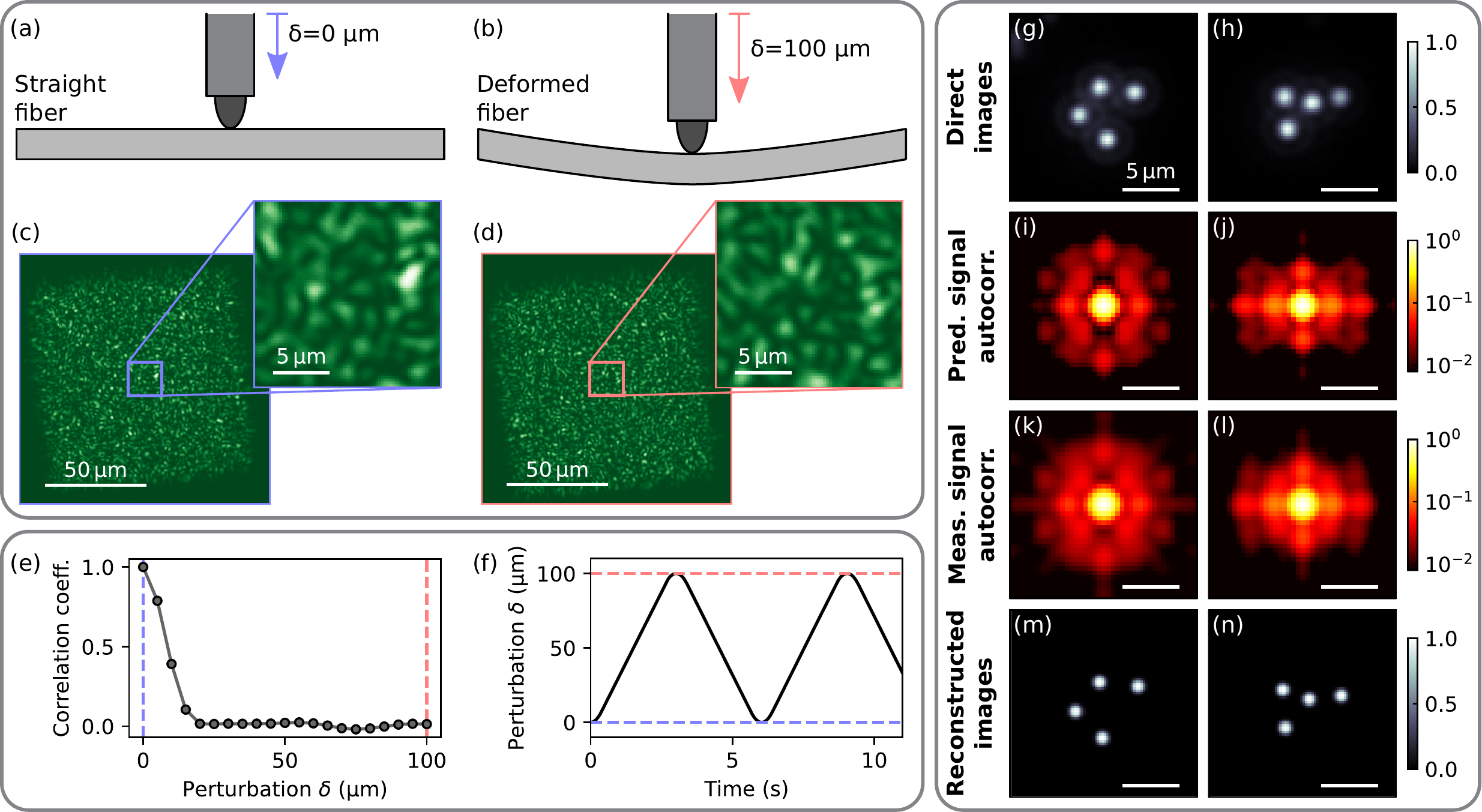}
	\end{center}
	\caption{Reconstructed images of fluorescent beads through a dynamically-perturbed square-core fiber. (a-b) Sketch of the deformation procedure used to dynamically perturb a $10.5$\,cm long fiber (sketch not to scale). (c-d) Transmitted speckle patterns for $\delta=0$\,\textmu m (straight fiber) and $\delta=100$\,\textmu m (deformed fiber), demonstrating that the applied perturbation leads to a decorrelation of the transmitted field. (e) Correlation coefficient as a function of the applied perturbation, taking as a reference the speckle pattern measured for $\delta=0$\,\textmu m. (f) Temporal dependence of the perturbation applied to the fiber during the experiments. This periodic perturbation is applied during the full acquisition time (approximately $4$ hours). (g-h) Direct fluorescence images of objects composed of fluorescent beads located at the output of the fiber. (i-j) Predicted signal autocorrelations calculated with the true position of the beads using \eq{eq_corr}. (k-l) Signal autocorrelations retrieved from experimental measurements (averaged over $N_{\mathrm{rep}}=40,000$ speckle illuminations). (m-n) Reconstructed images of the beads. Despite the dynamic perturbation applied to the fiber, the position of the beads is accurately retrieved.}
	\label{fig_perturbations}
\end{figure*}

\subsection*{Imaging through a dynamically-perturbed fiber}

Endoscopic imaging techniques based on transmission matrix measurements are known to be very sensitive to external perturbations such as vibrations and fiber bending. In contrast, our approach based on speckle correlations is intrinsically robust to such perturbations. To experimentally demonstrate this crucial advantage, we used a $10.5$\,cm long fiber, which can be more easily deformed as compared to the $3$\,cm long fiber. To ensure that this longer fiber length does not affect the efficiency of the method, we can compare the intensity correlation function $C_I (\Delta \bm{r}_a,\Delta \bm{r}_b)$ measured for the $3$\,cm long fiber and for the $10.5$\,cm long fiber. These two correlation functions are very similar (see Supplementary Section~3), evidencing that the kaleidoscopic memory effect is robust in this range of fiber lengths. 

To verify that our approach is robust to external perturbations, we deformed the fiber by pushing it at mid-length using a rod controlled by a motorized stage (\fig{fig_perturbations}a,b). Such a perturbation strongly modifies the propagation of light through the fiber, as can be verified by generating a random speckle pattern at the fiber input and by measuring output speckle patterns for different displacements $\delta$ of the rod. Speckle patterns measured for $\delta=0$\,\textmu m and $\delta=100$\,\textmu m are strongly different (\fig{fig_perturbations}c,d), due to a decorrelation of the measured patterns as well as to a transverse shift of the fiber (see Supplementary Section~8). To quantitatively analyze this perturbation, we calculated the correlation coefficient of the speckle patterns as a function of the displacement $\delta$, taking as a reference the speckle pattern measured for $\delta=0$\,\textmu m. The value of the correlation coefficient decreases from one to zero for a displacement of the rod of approximately $20$\,\textmu m (\fig{fig_perturbations}e), evidencing that the transmission matrix of the imaging system is completely modified by a displacement $\delta \geq 20$\,\textmu m. In contrast, the intensity correlation function $C_I (\Delta \bm{r}_a,\Delta \bm{r}_b)$ remains identical before and after a displacement of $100$\,\textmu m (see Supplementary Section~3), which demonstrates the robustness of the kaleidoscopic memory effect to such perturbations.

In our experiments, to reproduce the dynamical aspect of the perturbations that typically occur when studying living organisms, we continuously modified the position of the rod perturbing the fiber by applying a periodic displacement over a range of $100$\,\textmu m (\fig{fig_perturbations}f). We show in Supplementary Movie~M1 the decorrelation of the speckle patterns induced by this periodic displacement of the rod. While the time period of the perturbation ($6$\,s) is much shorter than the total acquisition time (approximately $4$\,hours), this does not affect the efficiency of our approach. Indeed, some stability is required when scanning a given input speckle pattern ($220$\,ms in our experiments), but the system can be perturbed between two different random realizations of the input speckle pattern. 

We used this dynamically-perturbed fiber to successively study two different objects, each composed of four fluorescent beads (\fig{fig_perturbations}g,h). The predicted signal autocorrelations (\fig{fig_perturbations}i,j) are again in excellent agreement with the measured signal autocorrelations (\fig{fig_perturbations}k,l), and images are faithfully reconstructed by the reconstruction algorithm (\fig{fig_perturbations}m,n). Note that similar results were also obtained by maintaining the fiber in a static position (see Supplementary Section~9), which confirms that the dynamical aspect of the applied perturbation does not significantly influence the efficiency of the method. As such, these results explicitly demonstrate that our approach is robust to fiber perturbations, far beyond what is achievable with transmission matrix measurements.

\section*{Discussion}

\begin{figure}[ht]
	\begin{center}
		\includegraphics[width=\linewidth]{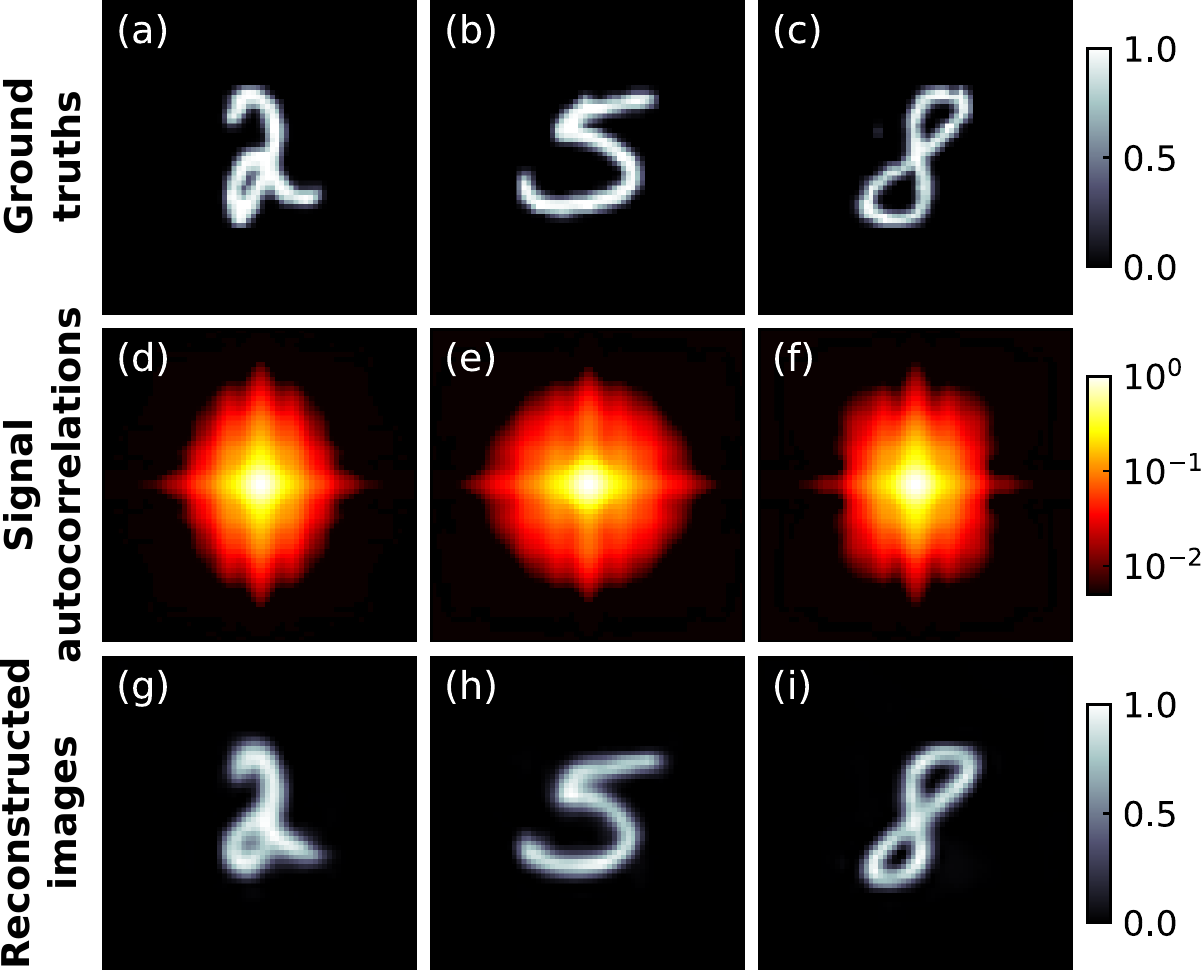}
	\end{center}
	\caption{Reconstructed images of handwritten digits in numerical simulations. (a-c)~Grayscale images from the MNIST database of handwritten digits~\cite{lecun_gradient-based_1998}, that are used as objects in our numerical simulations. (d-f)~Signal autocorrelations of these objects calculated using \eq{eq_corr}. (g-i)~Images reconstructed by an artificial neural network, demonstrating that the inverse problem underlying our approach can be successfully solved even in the case of continuous objects.}
	\label{fig_ann}
\end{figure}

To further investigate the possibility to image more complicated objects using the same approach, we performed a complementary study based on numerical simulations. For this purpose, we used as objects grayscale images from the MNIST database of handwritten digits (\fig{fig_ann}a-c). Using \eq{eq_corr}, we then numerically generated the associated signal autocorrelations (\fig{fig_ann}d-f), based on the intensity correlation function $C_I(\Delta \bm{r}_a,\Delta \bm{r}_b)$ that was measured in the experiment. This constitutes a direct approach to model the signal autocorrelations that would be measured with our optical setup, taking into account the finite numerical aperture of the fiber and the limited range of the memory effect, but however without describing the influence of statistical fluctuations. We then trained two different artificial neural networks in parallel to solve the inverse problem from these signal autocorrelations (see Methods), one for classification (a DenseNet~\cite{huang_densely_2017}) and one for image reconstruction (a UNeXt~\cite{valanarasu_unext_2022}). Testing these neural networks on unseen objects, we obtain a classification accuracy of $91$\% using the DenseNet, and images are reconstructed with a high fidelity by the UNeXt (\fig{fig_ann}g-i, see also Supplementary Section~8), with an average structural similarity between ground truths and reconstructed images of $0.89$. This demonstrates that the inverse problem underlying our approach can be successfully solved not only in the case of a few point-like objects but also in the case of more complicated, continuous objects.

For many applications, the acquisition time is an important parameter that should be minimized. While the acquisition time was approximately $4$\,hours in our proof-of-principle experiments, there exists several opportunities to increase the speed of such measurements. For instance, in our work, input fields are translated with a step size of $0.4$\,\textmu m, while the resolution limit of the system is $\lambda/(2 \mathrm{NA})\simeq 1.2$\,\textmu m; one could thus use a larger step size without impacting the resolution of reconstructed images. Other ideas could also be implemented to reduce the acquisition time, such as working with a more sensitive photomultiplier, brighter fluorescent beads, and a DMD with a larger on-board memory (see Methods). In addition, the number of different input fields needed to reconstruct the objects could also be strongly decreased. Indeed, while we worked with $40,000$ realizations to minimize the apparition of fluctuation-related artifact, the quality of reconstructed images is already relatively good with $3,000$ realizations (see Supplementary Section~S5). Furthermore, it might be envisioned that better reconstruction algorithms (e.g. based on artificial neural networks) could be used to reduce the influence of statistical fluctuations on the quality of the reconstructed images. 

Different strategies can also be envisioned in order to tackle limitations imposed by the restricted field of view. In our experiments, we created a single area of interest by photobleaching all other beads within the field of view of the fiber. However, if the fiber can collect light from several objects of interests that are not in the range of the memory effect, the resulting signal autocorrelation is the (incoherent) sum of the signal autocorrelations associated with each object. In this case, it might then be possible to unscramble these contributions and retrieve an image of each object, even though the relative position between these objects would remain unknown. An alternative option would be to deposit an opaque coating at the fiber output, in order to reduce the field of view of the fiber down to the area covered by the memory effect. In parallel, there exists several opportunities to extend the field of view of the approach, in the perspective of studying larger objects. A possible way to extend the field of view consists in increasing the range of the memory effect by optimizing the optical properties of the fiber, so that boundary conditions at the core-cladding interface are closer to those of a perfect mirror. It might also be possible to combine the spatial information available via the kaleidoscopic memory effect with the deterministic information provided by the knowledge of the eigenbasis of the fiber \cite{ploschner_seeing_2015} in order to extent the field of view. Finally, our approach could also benefit from the use of matrix factorization algorithms, not only to reconstruct images beyond the range of the memory effect~\cite{zhu_large_2022} but also to probe the dynamics of fluorescent objects~\cite{moretti_readout_2020}.

\section*{Conclusion}

To conclude, we introduced an approach to reconstruct images through a multimode optical fiber based on speckle correlations, without any prior information on the fiber. This approach takes advantage of symmetries in square-core fibers, which induce a kaleidoscopic memory effect that can be exploited to reconstruct images through the fiber. As an illustration, we reconstructed images of samples composed of several fluorescent beads, as being relevant e.g. for applications involving fluorescent emitters as functional indicators~\cite{weisenburger_guide_2018,turtaev_high-fidelity_2018}. Moreover, we demonstrated that our approach is robust to dynamic fiber perturbations. Finally, using numerical simulations, we evidenced that the inverse problem underlying the approach can be successfully solved even in the case of more complex objects. We anticipate that better strategies to solve this inverse problem will emerge from the recently-established fields of compressed sensing~\cite{eldar_compressed_2012} and deep learning~\cite{barbastathis_use_2019}, e.g. by reducing the influence of statistical fluctuations and by finding adequate sparsity constraints. Furthermore, we highlight that our method is generally applicable not only to fluorescent objects but also to any sample that generates a signal in response to light, such as photo-acoustic emission from optical absorbers~\cite{wang_practical_2016,caravaca-aguirre_hybrid_2019,zhou_photoacoustic_2020} or second-harmonic generation from non-linear materials~\cite{campagnola_second_2011,cifuentes_polarization-resolved_2021}.

\section*{Acknowledgments}

The authors thank M.\ Balland, A.\ Carron, A.\ Goetschy and S.\ Mezil for insightful discussions. This work was supported by the H2020 European Research Council (grant 681514-COHERENCE) and by a Marie Skłodowska Curie Individual Fellowship (grant 750420-DARWIN).

\section*{Data and Software Availability}

Raw experimental data and Python scripts are available (DOI: \href{https://doi.org/10.57745/B6PSX0}{https://doi.org/10.57745/B6PSX0}).

\section*{Materials and Methods}

	\subsection*{Sample preparation}
	
	Fluorescent objects are composed of latex microspheres (ThermoFisher Scientific, red FluoSpheres, diameter $1.0$\,\textmu m) dispersed on a glass coverslip. A solution of Poly-L-lysine (Sigma-Aldrich, 0.1\,\% in H$_2$O) is first deposited on a clean coverslip in order to fix the microspheres. A solution of fluorescent beads diluted in water is then deposited on the coverslip. Using this procedure, fluorescent beads are randomly dispersed on the sample. In order to detect only the fluorescence of the few beads of interest, we selectively photobleach all other beads within the field of view of the fiber (area of $100$\,\textmu m\,$\times$\,$100$\,\textmu m) by successively focusing coherent light from a continuous-wave laser (Cobolt 08-DPL, 532\,nm, 10\,mW after attenuation by a neutral density filter) on the beads using a $\times20$ objective (Mitutoyo Plan Apo SL 20X/0.28).
	
	\subsection*{Experimental setup}
	
	The optical setup that we built to image fluorescent objects through a multimode fiber is represented in Supplementary Section~1. Coherent light is generated by a continuous-wave solid-state laser (Cobolt 08-DPL, $532$\,nm). Light is injected in a single-mode polarization-maintaining fiber and outcoupled using a collimator (Schäfter+Kirchhoff 60FC-L-4-M75-01). Light then passes through a linear polarizer to ensure that it is horizontally polarized, before being reflected by a digital micromirror device (Vialux Superspeed V-7001). Random speckle patterns are generated and translated in the plane of the fiber using Lee holography \cite{lee_binary_1974}. This technique is implemented using a 4f system composed of a $300$\,mm lens (L1) and a $150$\,mm lens (L2). In the focal plane between these lenses, an iris selects the first diffraction order of the grating displayed by the DMD. A dichroic mirror (Chroma ZT532rdc) reflects the light towards a $\times20$ objective (Nikon CF Plan 20X/0.35 EPI SLWD), and a square-core multimode fiber (CeramOptec, core section $100$\,\textmu m\,$\times$\,$100$\,\textmu m, cladding diameter $123$\,\textmu m, numerical aperture $0.22$) is placed in its focal plane. The input speckle patterns generated using this procedure are characterized by a numerical aperture of $0.22$ (to match that of the fiber) and a spatial extent of $92$\,\textmu m\,$\times$\,$92$\,\textmu m. This allows us to translate the patterns over an area of $8$\,\textmu m\,$\times$\,$8$\,\textmu m without illuminating the fiber cladding. Note that, to avoid autofluorescence of the fiber coating, we removed it using a solution of trichloromethane. 
	
	The sample, located approximately $20$\,\textmu m away from the fiber output, is thus illuminated by light coming from the fiber. Fluorescence light is then collected by the same fiber and, after passing through the dichroic mirror, it is spatially filtered using a 4f system composed of a $200$\,mm lens (L3) and a $100$\,mm lens (L4), with an iris located in the focal plane in-between these lenses. This iris is used to block unwanted light coming from outside of the fiber core. Light is then spectrally filtered using two successive fluorescence filters (Thorlabs NF533-17 and Chroma ET577LP) and focused using a $150$\,mm lens onto a photomultiplier tube module (Hammamatsu H7422P). The measured analog signal passes a low-noise current amplifier (Stanford Research Systems SR570), which applies a low-pass filter to the signal (-6\,dB cutoff frequency: 10\,kHz) in order to improve the signal-to-noise ratio. Finally, this analog signal is converted into a digital signal by an acquisition board (National Instruments PCIe-6321).
	
	Dynamic deformations are applied to the fiber by holding the fiber at both extremities and by pushing it at mid-length using a small rod (diameter $2$\,mm), along the $y$ direction (corresponding to the vertical direction in our experiments). The position of the rod is controlled using a motorized stage (PI M-230.25) connected to a different computer, ensuring that measurements and perturbations are performed in an asynchronous manner.
	
	\subsection*{Acquisition procedure}
	
	To minimize the acquisition time, we first pre-calculate the patterns that will be displayed by the DMD. These patterns are split into packets of $79$ realizations of input speckles, each containing $21$\,$\times$\,$21$ patterns which are translated versions of the same speckle. These files are then stored on a solid-state drive (SSD) in a binary format. During an acquisition, each file is loaded on the random access memory (RAM) of the computer (loading time, 2.4\,s), then transferred into the internal memory of the DMD (transfer time, 8.7\,s), and finally displayed by the DMD running at a rate of 2\,kHz (display time, 17.4\,s). Thus, overall, measuring data for $40,000$ realizations takes approximately $4$ hours. Note that, while our DMD could be operated at a rate of up to 23\,kHz, working at $2$\,kHz allows us to improve the signal-to-noise ratio by applying a low-pass filter (-6\,dB cutoff frequency: 10\,kHz) to the measured signal. Finally, to obtain direct images of the sample, we used a $\times20$ objective (Mitutoyo Plan Apo SL 20X/0.28) located on the other side of the sample, along with a 200\,mm lens, a fluorescence filter (Chroma ET590/50m) and a complementary metal oxide semiconductor (CMOS) camera (Basler acA1300-200um). Direct images were then obtained by averaging the measured images over random illumination patterns coming from the fiber.

	\subsection*{Numerical simulations}
	
	In the numerical simulations, we use as objects $28\times28$ grayscale images from the MNIST database of handwritten digits~\cite{lecun_gradient-based_1998}. For consistency with experimental measurements of the intensity correlation function $C_I(\Delta \bm{r}_a,\Delta \bm{r}_b)$, these images are re-scaled to produce $64\times64$ images, using a magnification factor so that the thickness of the lines forming the digits approximately matches the apparent width of a speckle grain. Hence, in physical units, the field of view of the images presented in \fig{fig_ann} would be $16.4$\,\textmu m\,$\times \,16.4$\,\textmu m (which is the actual field of view of the images presented in \fig{fig_results}). We then numerically calculate the signal autocorrelations for each object using \eq{eq_corr}, using for the intensity correlation function the one measured on the $3$\,cm long fiber (see Supplementary Section~3). Finally, signal autocorrelations are re-sampled to produced $64\times64$ images, to be used at the input of the artificial neural network. 
	
	The deep learning model used for image reconstruction is a UNeXt~\cite{valanarasu_unext_2022}, which is a modified version of the U-Net~\cite{ronneberger_u-net_2015} and the ResUNet~\cite{diakogiannis_resunet-_2020} architectures. Thus, our convolutional neural network is composed of two symmetric networks, an encoder and a decoder, each composed of 4 convolutional blocks. The encoder extracts complex features to produce a representation of the input in a latent space while the decoder produces output images with the same resolution as input ones. Context information is propagated from the encoder to the decoder through 4 skip connections (one for each convolutional block), which provide local information to the global information. As a specificity of the UNeXt architecture, our neural network also includes a convolutional multilayer perception block, in order to produce a better representation of the data. We train this neural network via a conventional scheme, using the Adam optimizer, a mean-squared-error loss function, a batch size of 32 and an early stopping procedure. We use the 60,000 examples of the MNIST database as follows: 55,800 examples for training, 3,000 examples for validation, and 1,200 examples for testing. Finally, the performances of the network is assessed using the average structural similarity (SSIM), which allows one to compare the visual similarity between two images. 
	
	For classification, we used a DenseNet \cite{huang_densely_2017}, a convolutional neural network with short connections between layers inside each convolutional block. Each layer obtains additional inputs from all preceding layers and passes on its own feature-maps to all subsequent layers. Our DenseNet is composed of 3 dense convolutional blocks, each of them composed of 6 layers. We train this neural network via a conventional scheme, using the Adam optimizer, a cross-entropy loss function, a batch size of 32 and an early stopping procedure. As for the UNeXt, we use the same 55,800 examples for training, 3,000 examples for validation, and 1,200 examples for testing.


\bibliography{references}


\onecolumngrid
\pagebreak
\beginsupplement
\begin{center}
	\textbf{\large Speckle-correlation imaging through a kaleidoscopic multimode fiber\\ \bigskip Supplementary information}
	
	\bigskip
	Dorian Bouchet,$^1$ Antonio M.\ Caravaca-Aguirre,$^1$ Guillaume\\Godefroy,$^{1,\,2}$ Philippe Moreau,$^1$ Ir\`ene Wang,$^1$ and Emmanuel Bossy$^1$\\ \vspace{0.15cm}
	\textit{\small $^\mathit{1}$Université Grenoble Alpes, CNRS, LIPhy, 38000 Grenoble, France}\\
	\textit{\small $^\mathit{2}$Université Grenoble Alpes, CEA, Leti, 38000 Grenoble, France}
\end{center}
\vspace{0.3cm}


\section{Schematic representation of the optical setup}

\begin{figure*}[ht]
	\begin{center}
		\includegraphics[width=0.6\linewidth]{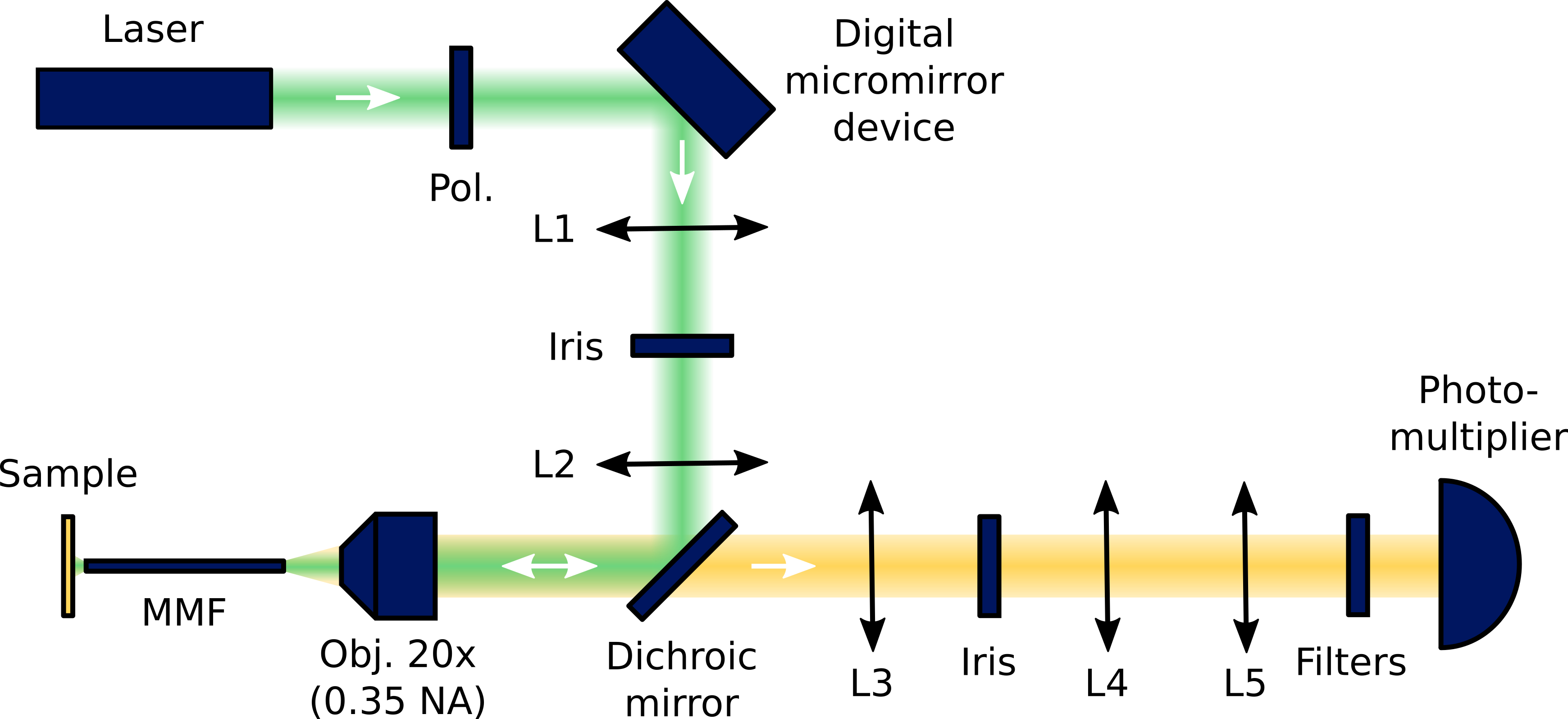}
	\end{center}
	\caption{\textbf{Optical setup.} Speckle patterns are generated using a digital micromirror device (DMD) and translated at the input of a square-core multimode fiber. A sample composed of fluorescent beads is placed at the output of the fiber. Fluorescence collected through the fiber passes a dichroic mirror and is detected by a single-channel photomultiplier. MMF: multimode fiber; Pol, linear polarizer; Obj, objective; NA, numerical aperture; L1 to L5, lenses with focal length 300\,mm (L1), 150\,mm (L2), 200\,mm (L3), 100\,mm (L4) and 150\,mm (L5).}
	\label{fig_setup}
\end{figure*}


\section{Calculation of the signal autocorrelation}

\subsection{General expression}

The density distribution of fluorescent emitters $O(\bm{r}_b)$ is related to the measured fluorescence signal $S(\bm{r}_a)$ by the following relation:
\begin{equation}
	S(\bm{r}_a) = \int O(\bm{r}_b) I(\bm{r}_a,\bm{r}_b) \, \de \bm{r}_b ,
	\label{eq_signal_raw}
\end{equation}
where $I(\bm{r}_a,\bm{r}_b)$ is the excitation intensity transmitted at a position $\bm{r}_b$ at the fiber output for a translation $\bm{r}_a$ of the speckle pattern at the fiber input. The statistical autocorrelation function of the fluorescence signal is defined as follows:
\begin{equation}
	C_S (\bm{r}_a,\bm{r}_a') = \frac{\langle S(\bm{r}_a) S(\bm{r}_a')\rangle - \langle S(\bm{r}_a) \rangle \langle S(\bm{r}_a')\rangle }{\sqrt{\langle S(\bm{r}_a)^2\rangle - \langle S(\bm{r}_a)\rangle^2} \times \sqrt{\langle S(\bm{r}_a')^2\rangle - \langle S(\bm{r}_a')\rangle^2} } ,
	\label{eq_cs}
\end{equation}
where $\langle \dots \rangle$ represents the average over different realizations of the speckle pattern at the fiber input. We also define the contrast of the signal $K(\bm{r}_a)$ as follows:
\begin{equation}
	K(\bm{r}_a) = \frac{\sqrt{\langle S (\bm{r}_a)^2 \rangle - \langle S (\bm{r}_a) \rangle^2 }}{\langle S (\bm{r}_a) \rangle } .
\end{equation}
Using this definition, \eq{eq_cs} reads
\begin{equation}
	C_S (\bm{r}_a,\bm{r}_a') = \frac{\langle S(\bm{r}_a) S(\bm{r}_a')\rangle - \langle S(\bm{r}_a) \rangle \langle S(\bm{r}_a') \rangle }{ K(\bm{r}_a) \langle S(\bm{r}_a)\rangle \times K(\bm{r}_a')\langle S(\bm{r}_a')\rangle } .
\end{equation}	
Using \eq{eq_signal_raw}, the average fluorescence signal can be expressed as follows:
\begin{equation}
	\langle S(\bm{r}_a) \rangle = \langle I(\bm{r}_a,\bm{r}_b) \rangle \int O (\bm{r}_b) \de \bm{r}_b .
	\label{eq_avg_signal}
\end{equation}
Moreover, the correlator $\langle S(\bm{r}_a) S(\bm{r}_a' )\rangle$ is given by
\begin{equation}
	\langle S(\bm{r}_a) S(\bm{r}_a ')\rangle = \iint O (\bm{r}_b) O (\bm{r}_b') \langle I(\bm{r}_a,\bm{r}_b) I(\bm{r}_a',\bm{r}_b') \rangle \de \bm{r}_b \de \bm{r}_b ' .
	\label{eq_corr_signal}
\end{equation}
Inserting \eqs{eq_avg_signal}{eq_corr_signal} into \eq{eq_cs} yields
\begin{equation}
	C_S (\bm{r}_a,\bm{r}_a') = \cfrac{\iint O (\bm{r}_b) O (\bm{r}_b') \left[ \langle I(\bm{r}_a,\bm{r}_b) I(\bm{r}_a',\bm{r}_b') \rangle - \langle I(\bm{r}_a,\bm{r}_b) \rangle \langle I(\bm{r}_a',\bm{r}_b') \rangle \right] \de \bm{r}_b \de \bm{r}_b '}{ K(\bm{r}_a)  K(\bm{r}_a') \left[ \int O (\bm{r}_b) \de \bm{r}_b \right]^2 \langle I(\bm{r}_a,\bm{r}_b) \rangle \langle I(\bm{r}_a',\bm{r}_b') \rangle } .
	\label{eq_cs2}
\end{equation}	
In this expression, we can recognize the definition of the statistical correlation function of the intensity: 
\begin{equation}
	C_I(\bm{r}_a,\bm{r}_b,\bm{r}_a',\bm{r}_b') = \frac{\langle I(\bm{r}_a,\bm{r}_b) I(\bm{r}_a',\bm{r}_b')\rangle - \langle I(\bm{r}_a,\bm{r}_b) \rangle \langle I(\bm{r}_a',\bm{r}_b') \rangle }{\langle I(\bm{r}_a,\bm{r}_b) \rangle \langle I(\bm{r}_a',\bm{r}_b') \rangle } .
	\label{def_CI}
\end{equation}
Note that this correlation function is normalized: the speckle is assumed to be fully developed, so that $\langle I(\bm{r}_a,\bm{r}_b) \rangle=\sqrt{\langle I(\bm{r}_a)^2\rangle - \langle I(\bm{r}_a)\rangle^2}$. Inserting \eq{def_CI} into \eq{eq_cs2} yields
\begin{equation}
	C_S (\bm{r}_a,\bm{r}_a') = \cfrac{\iint O (\bm{r}_b) O (\bm{r}_b') C_I(\bm{r}_a,\bm{r}_b,\bm{r}_a',\bm{r}_b') \de \bm{r}_b \de \bm{r}_b '}{ K(\bm{r}_a)  K(\bm{r}_a') \left[ \int O (\bm{r}_b) \de \bm{r}_b \right]^2} .
\end{equation}	
Shift-shift memory effects~\cite{osnabrugge_generalized_2017_2} (including the translational memory effect observed in square-core fibers) are described by an intensity correlation function $C_I(\bm{r}_a,\bm{r}_b,\bm{r}_a',\bm{r}_b')=C_I(\Delta \bm{r}_a,\Delta \bm{r}_b)$, where $\Delta \bm{r}_a=\bm{r}_a'-\bm{r}_a$ and $\Delta \bm{r}_b = \bm{r}_b' - \bm{r}_b$. In addition, $ \langle I(\bm{r}_a,\bm{r}_b) \rangle$ and $ \langle I(\bm{r}_a,\bm{r}_b)^2 \rangle$ are independent of $\bm{r}_a$ and $\bm{r}_b$, implying that the signal contrast $ K$ is also independent of $\bm{r}_a$. This yields
\begin{equation}
	C_S (\Delta \bm{r}_a) = \cfrac{\iint O (\bm{r}_b) O (\bm{r}_b+\Delta \bm{r}_b) C_I(\Delta \bm{r}_a,\Delta \bm{r}_b) \de \bm{r}_b \de \Delta \bm{r}_b }{ K^2 \left[ \int O (\bm{r}_b) \de \bm{r}_b \right]^2} .
	\label{eq_cs3}
\end{equation}	
In this expression, we can recognize the spatial autocorrelation function of the object: 
\begin{equation}
	C_O(\Delta \bm{r}_b) = \frac{\int O (\bm{r}_b) O (\bm{r}_b + \Delta \bm{r}_b) \de \bm{r}_b }{\left[ \int O (\bm{r}_b) \de \bm{r}_b \right]^2} .
	\label{def_CO}
\end{equation}
Inserting \eq{def_CO} into \eq{eq_cs3} yields
\begin{equation}
	C_S (\Delta \bm{r}_a) = K^{-2} \int C_O(\Delta \bm{r}_b) C_I(\Delta \bm{r}_a,\Delta \bm{r}_b) \de \Delta \bm{r}_b .
	\label{CS_general}
\end{equation}
This expression, given as Eq.~(2) in the manuscript, relates the autocorrelation function of the fluorescence signal $C_S (\Delta \bm{r}_a)$, the object autocorrelation function $C_O(\Delta \bm{r}_b)$ and the intensity correlation function of the coherent excitation field $C_I(\Delta \bm{r}_a,\Delta \bm{r}_b)$.

\subsection{Limiting case}

For an ideal model of square-core fibers with an infinite-range memory effect, the intensity correlation function $C_I(\Delta \bm{r}_a,\Delta \bm{r}_b)$ is given by~\cite{caravaca-aguirre_optical_2021_2}
\begin{equation}
	C_I(\Delta \bm{r}_a,\Delta \bm{r}_b) = \frac{1}{16} \sum_{j=1}^4 \sum_{k=1}^4 C_j (\Delta \bm{r}_a,\Delta \bm{r}_b) C_k (\Delta \bm{r}_a,\Delta \bm{r}_b).
	\label{CI_ideal}
\end{equation}
In this expression, the terms $C_j$ are field correlation functions defined as follows:
\begin{equation}
	C_j (\Delta \bm{r}_a,\Delta \bm{r}_b) = \cfrac{2 J_1 (k_0 \, \mathrm{NA} \, \sqrt{(\Delta x_{b}-\xi_{x,j}\Delta x_{a})^2+(\Delta y_{b}-\xi_{y,j}\Delta y_{a})^2})}{k_0 \, \mathrm{NA} \, \sqrt{(\Delta x_{b}-\xi_{x,j}\Delta x_{a})^2+(\Delta y_{b}-\xi_{y,j}\Delta y_{a})^2}} ,
	\label{Cj}
\end{equation}
where $J_1$ is the first-order Bessel function of the first kind, $k_0=2 \pi / \lambda$ is the wavenumber, $\mathrm{NA}$ is the numerical aperture of the fiber, $\xi_{x,1}=\xi_{x,3}=\xi_{y,1}=\xi_{y,2}=1$ and $\xi_{x,2}=\xi_{x,4}=\xi_{y,3}=\xi_{y,4}=-1$. Note that field correlation functions defined by \eq{Cj} are composed of a single peak. These functions constitute the building blocks of the intensity correlation function given by \eq{CI_ideal}, which is characterized by four peaks that translate with $\Delta \bm{r}_a$ and that coherently overlap by pair when $\Delta x_a=0$ or $\Delta y_a=0$.

Assuming that the correlation functions $C_j (\Delta \bm{r}_a,\Delta \bm{r}_b)$ are infinitely sharp (i.e. assuming that the size of the speckle grain is small as compared to the smallest features of the object), we can find approximate expressions for the product $C_j (\Delta \bm{r}_a,\Delta \bm{r}_b) C_k (\Delta \bm{r}_a,\Delta \bm{r}_b)$. When $\Delta x_{a} (\xi_{x,j}-\xi_{x,k}) =0$ and $\Delta y_{a}(\xi_{y,j}-\xi_{y,k}) =0 $, we can write
\begin{equation}
	C_j (\Delta \bm{r}_a,\Delta \bm{r}_b) C_k (\Delta \bm{r}_a,\Delta \bm{r}_b) = A \delta (\Delta x_{b}-\xi_{x,j}\Delta x_{a}) \delta (\Delta y_{b}-\xi_{y,j}\Delta y_{a}),
\end{equation}
where $\delta$ denotes the Dirac delta function and $A=4 \pi / (k_0 \mathrm{NA})^2$ is the area covered by a speckle grain. In contrast, when $\Delta x_{a} (\xi_{x,j}-\xi_{x,k}) \neq 0 $ or $\Delta y_{a} (\xi_{y,j}-\xi_{y,k}) \neq 0 $, we can write
\begin{equation}
	C_j (\Delta \bm{r}_a,\Delta \bm{r}_b) C_k (\Delta \bm{r}_a,\Delta \bm{r}_b) = 0. 
\end{equation}
Using these approximations to calculate the intensity correlation function given by \eq{CI_ideal}, the signal autocorrelation expressed by \eq{CS_general} can be simplified as follows:
\begin{itemize}
	\setlength{\itemsep}{0pt}
	\setlength{\parskip}{-1pt}
	\setlength{\parsep}{0pt}
	\item when $\Delta x_a \neq 0 $ and $\Delta y_a \neq 0 $, we have 
	\begin{equation}
		C_S (\Delta \bm{r}_a)= \frac{A}{16 K^2} \left[ C_O (\Delta x_a ,\Delta y_a) + C_O (-\Delta x_a ,\Delta y_a)+C_O (\Delta x_a ,-\Delta y_a)+C_O (-\Delta x_a ,-\Delta y_a) \right];
	\end{equation}
	\item when $\Delta x_a = 0 $ and $\Delta y_a \neq 0 $, we have 
	\begin{equation}
		C_S (\Delta \bm{r}_a)= \frac{A}{4 K^2} \left[ C_O (\Delta x_a=0 ,\Delta y_a) + C_O (\Delta x_a=0 ,-\Delta y_a) \right];
	\end{equation}
	\item when $\Delta x_a \neq 0 $ and $\Delta y_a = 0 $, we have 
	\begin{equation}
		C_S (\Delta \bm{r}_a)= \frac{A}{4 K^2} \left[ C_O (\Delta x_a ,\Delta y_a=0) + C_O (-\Delta x_a ,\Delta y_a=0) \right];
	\end{equation}
	\item when $\Delta x_a = 0 $ and $\Delta y_a = 0 $, we have 
	\begin{equation}
		C_S (\Delta \bm{r}_a)= \frac{A}{K^2} C_O (\Delta x_a=0 ,\Delta y_a=0) .
	\end{equation}
\end{itemize}
Since $C_O$ is an even function, we can write $C_O (\Delta x_a,\Delta y_a)=C_O (-\Delta x_a,-\Delta y_a)$ and $C_O (\Delta x_a,-\Delta y_a)=C_O (-\Delta x_a,\Delta y_a)$. Defining $C_O^{\mathrm{sym}} (\Delta \bm{r}_a) = [C_O (\Delta x_a,\Delta y_a)+C_O (\Delta x_a,-\Delta y_a)]/2$ as the symmetrized version of the object autocorrelation, we obtain
\begin{equation}
	C_S (\Delta \bm{r}_a) = \frac{A}{K^2} \times 
	\begin{cases}
		C_O^{\mathrm{sym}} (\Delta \bm{r}_a)/4 \quad & \mathrm{if} \; \Delta x_{a} \neq 0 \; \mathrm{and} \; \Delta y_{a} \neq 0 ,\\
		C_O^{\mathrm{sym}} (\Delta \bm{r}_a)/2 \quad & \mathrm{if} \; \Delta x_{a} = 0 \; \mathrm{and} \; \Delta y_{a} \neq 0 ,\\
		C_O^{\mathrm{sym}} (\Delta \bm{r}_a)/2 \quad & \mathrm{if} \; \Delta x_{a} \neq 0 \; \mathrm{and} \; \Delta y_{a} = 0 ,\\
		C_O^{\mathrm{sym}} (\Delta \bm{r}_a) \quad & \mathrm{if} \; \Delta x_{a} = 0 \; \mathrm{and} \; \Delta y_{a} = 0 . \\
	\end{cases}
\end{equation}
This expression shows that, in the limiting case of an infinite-range memory effect with a speckle grain size approaching zero, the signal autocorrelation can be written as $C_S (\Delta \bm{r}_a) \propto w(\Delta \bm{r}_a) C_O^{\mathrm{sym}} (\Delta \bm{r}_a)$, where $w$ is a weight function that is equal to $1$ if $\Delta x_a=0$ and $\Delta y_a=0$, to $1/2$ if either $\Delta x_a = 0$ or $\Delta y_a = 0$, and to $1/4$ otherwise.


\section{Intensity correlation function of square-core multimode fibers}

\label{sec_CI}

\begin{figure}[!b]
	\begin{center}
		\includegraphics[width=\linewidth]{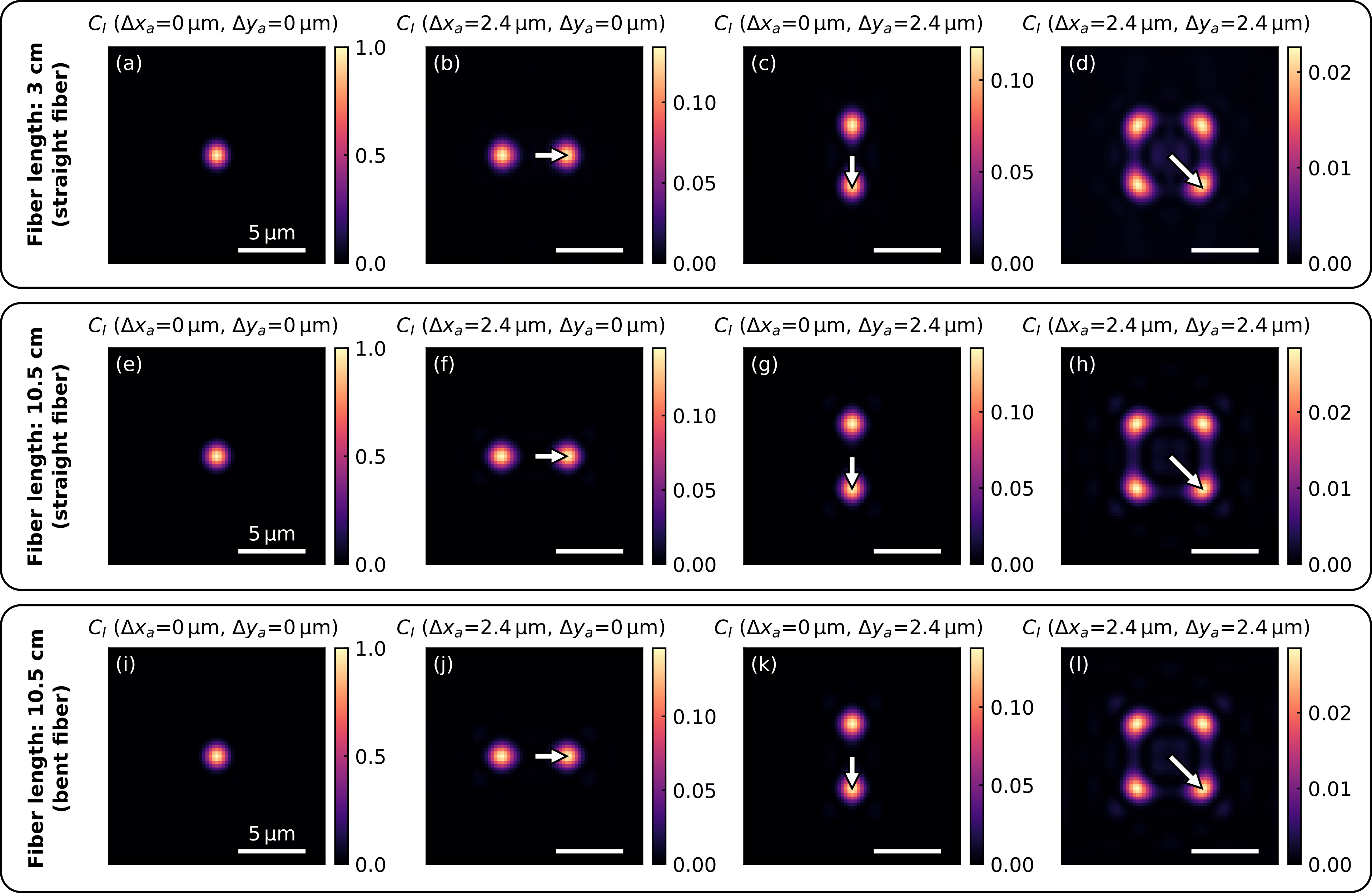}
	\end{center}
	\caption{(a-d) Intensity correlation function of the excitation field $C_I (\Delta \bm{r}_a, \Delta \bm{r}_b)$ for the $3$\,cm long fiber, held in a straight position. The correlation function is represented as a function of $\Delta \bm{r}_b$ (shift at the fiber output) for four different values of $\Delta \bm{r}_a$ (shift at the fiber input, represented by the white arrows on the figures). (e-h) Analogous to (a-d) for the $10.5$\,cm long fiber, held in a straight position. The range of the memory effect is not visibly affected by the longer length of the fiber (the observed range is even slightly larger). (i-l) Analogous to (a-d) for the $10.5$\,cm long fiber, held deformed. While this deformation fully decorrelates output speckle patterns, the range of the memory effect is not visibly affected by the fiber bending. Note that there is a twist of $5.5^\circ$ between the input and the output planes of the $10.5$\,cm long fiber; this angle is however not visible in the figures, as the coordinates $\Delta \bm{r}_b$ are defined in the rotated frame.}
	\label{fig_CI}
\end{figure}

In the experiment, due to the limited range of the kaleidoscopic memory effect, the intensity correlation function $C_I(\Delta \bm{r}_a,\Delta \bm{r}_b)$ differs from the one predicted by the ideal model [see \eqs{CI_ideal}{Cj}]. To experimentally measure this correlation function, we imaged the intensity of the coherent field at the fiber output using a $\times20$ objective (Mitutoyo Plan Apo SL 20X/0.28) along with a 200\,mm lens and a CMOS camera (Basler acA1300-200um). We generated $10,000$ random realizations of the input field, we translated them with the DMD ($\bm{r}_a$ covers a $8$\,\textmu m\,$\times$\,$8$\,\textmu m area), and we recorded the intensity measured by the camera ($\bm{r}_b$ covers a $14.6$\,\textmu m\,$\times$\,$14.6$\,\textmu m area). The normalized intensity correlation function $C_I(\Delta \bm{r}_a,\Delta \bm{r}_b)$ is then calculated from its definition [see \eq{def_CI}], averaging over all possible spatial pairs and all realizations of the input field. Finally, we take advantage of known symmetries and average the calculated normalized intensity correlation function with its left-right flipped version and its up-down flipped version, both on the input side and on the output side. As a result, we obtain the correlation function shown in \fig{fig_CI}a-d in the case of the $3$\,cm long fiber (also shown in Fig.~2d-g of the manuscript), and the correlation function shown in \fig{fig_CI}e-h in the case of the $10.5$\,cm long fiber. These correlation functions are very similar, indicating that the kaleidoscopic memory effect is robust in this range of fiber lengths and fiber bending. Finally, after perturbing the $10.5$\,cm long fiber by applying a displacement $\delta =100$\,\textmu m at mid-length, we obtain the correlation function shown in \fig{fig_CI}i-l. While this perturbation leads to a full decorrelation of the output speckle patterns (see Supplementary Section~\ref{sec_decorrelation}), the intensity correlation function $C_I(\Delta \bm{r}_a,\Delta \bm{r}_b)$ remains the same before and after the deformation, demonstrating that the memory effect is robust to such a perturbation. Note that, in the case of the $10.5$\,cm long fiber, the fiber was naturally twisted by an angle of $5.5^\circ$. In such a case, the memory effect follows the axes of the fiber, without any visible reduction in the range covered by the effect.


\section{Inherent ambiguities in the inverse problem}

\label{sec_ambiguities}

\begin{figure}[ht]
	\begin{center}
		\includegraphics[width=0.65\linewidth]{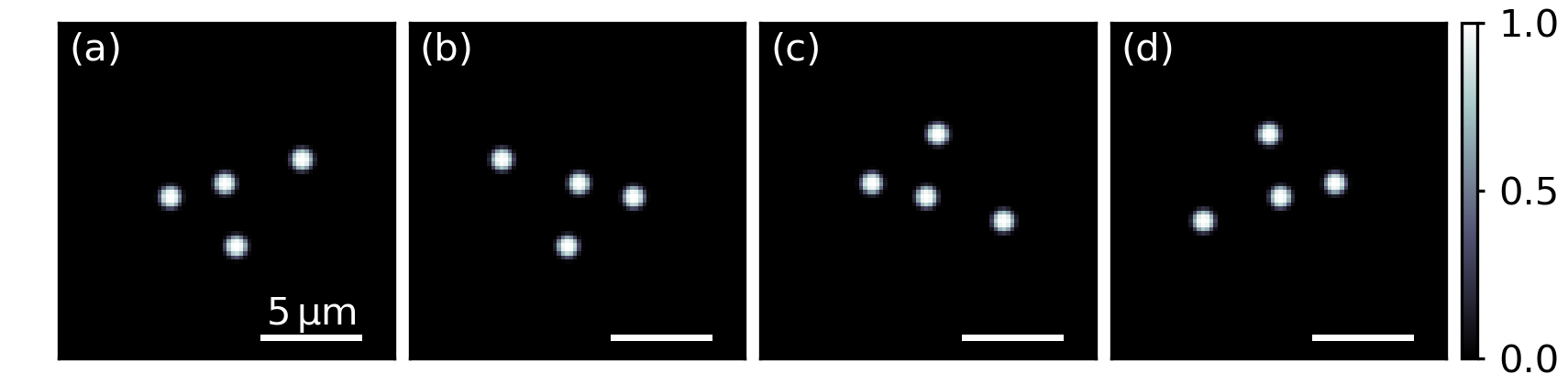}
	\end{center}
	\caption{Ambiguities in the inverse problem. These four different objects, that are flipped versions of each other, lead to the same symmetrized autocorrelation.}
	\label{fig_ambiguity}
\end{figure}

There are several possible solutions when trying to retrieve $O( \bm{r}_b)$ from $C_O^{\mathrm{sym}} (\Delta \bm{r}_b)$. First, as in usual inverse autocorrelation problems, any shift of the object in the transverse plane leads to the same symmetrized autocorrelation function. Moreover, since $C_O^{\mathrm{sym}} (\Delta x_b, \Delta y_b) = C_O^{\mathrm{sym}} (- \Delta x_b,\Delta y_b) = C_O^{\mathrm{sym}} (\Delta x_b,-\Delta y_b) = C_O^{\mathrm{sym}} (- \Delta x_b,- \Delta y_b) $, there exists four different objects that lead to the same symmetrized autocorrelation. These four objects are flipped versions of each other, as illustrated in \fig{fig_ambiguity}. Note that a similar ambiguity also exists in usual inverse autocorrelation problems~\cite{bertolotti_non-invasive_2012_2}: since $C_O (\Delta x_b, \Delta y_b) = C_O (- \Delta x_b,- \Delta y_b) $, there exists two different objects that lead to the same autocorrelation (the two objects that are represented in \fig{fig_ambiguity}a and \fig{fig_ambiguity}d).

\section{Influence of statistical fluctuations on the reconstructed images}

\begin{figure}[ht]
	\begin{center}
		\includegraphics[width=\linewidth]{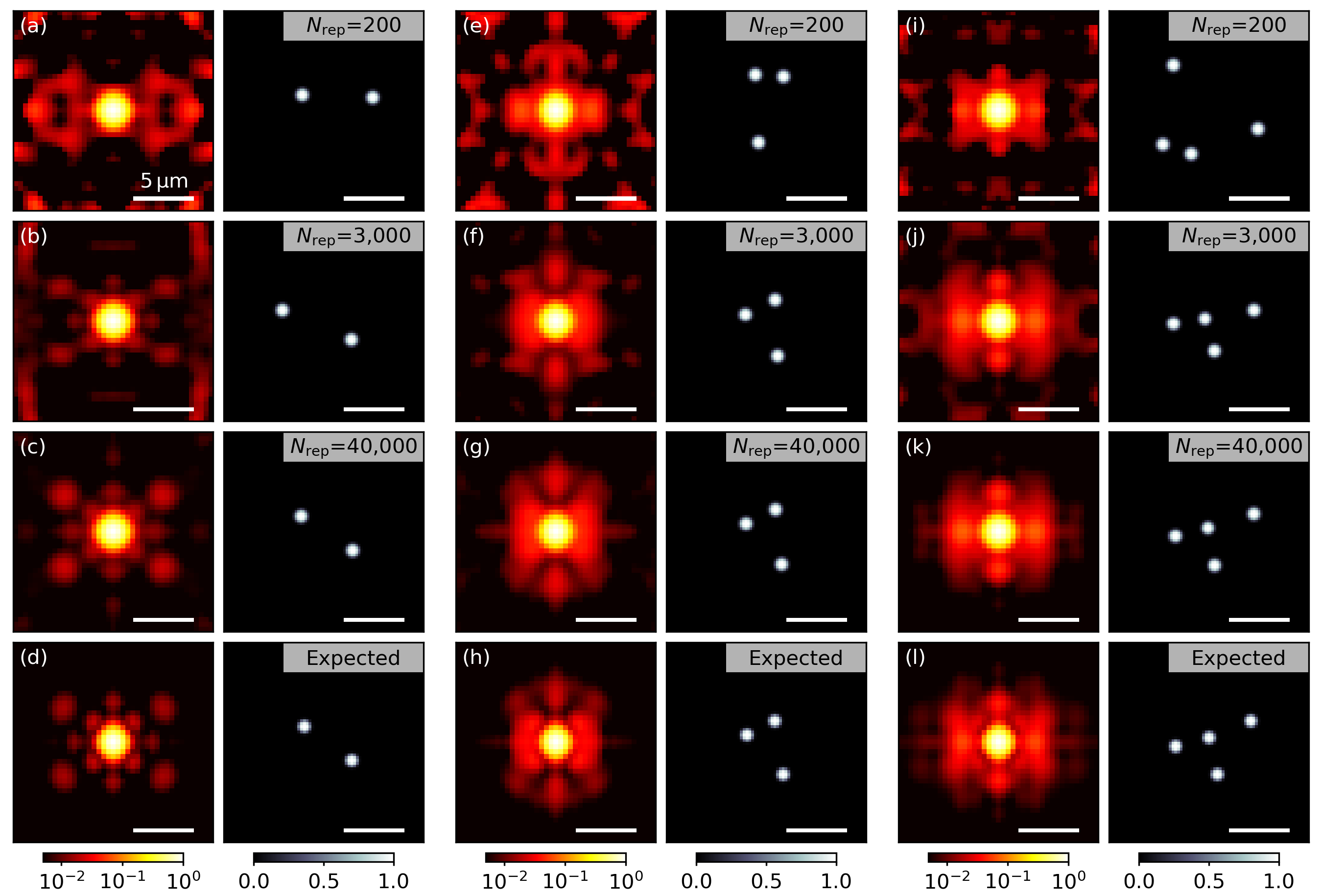}
	\end{center}
	\caption{Signal autocorrelations retrieved from experimental measurements and reconstructed images for different numbers of realizations of the input field. (a-c) Signal autocorrelation (top) and reconstructed images (bottom) for the sample composed of two beads, for $N_{\mathrm{rep}}=200$ realizations (a), $N_{\mathrm{rep}}=3,000$ (b) and $N_{\mathrm{rep}}=40,000$ (c). (d) Theoretically-predicted signal autocorrelation (left) along with the true position of the beads (right). (d-h) Analogous to (a-d) for the object composed of three fluorescent beads. (i-l) Analogous to (a-d) for the object composed of four fluorescent beads.}
	\label{fig_averaging}
\end{figure}

In order to minimize the influence of statistical fluctuations on the reconstructed images, the measured signal autocorrelation is averaged over random realizations of the input field. In the manuscript, we presented data obtained with $N_{\mathrm{rep}}=40,000$, yielding a signal correlation with no visible artifacts but requiring a significant acquisition time (4\,hours in total). Nevertheless, information about the object is still available even for much lower numbers of realizations. To illustrate this, we compare in \fig{fig_averaging} signal autocorrelations and reconstructed images for different numbers of realizations of the input field. 

\begin{itemize}
	\item For $N_{\mathrm{rep}}=200$ (acquisition time of 1\,min 12\,s), the shape of the objects is not faithfully reconstructed, but some distinctive features already appear on the signal autocorrelations (\fig{fig_averaging}a,e,i). 
	\item For $N_{\mathrm{rep}}=3,000$ (acquisition time of 18\,min), the shape of the objects can already be recognized, but a few beads are slightly mislocalized. In this case, signal autocorrelations strongly resemble the theoretical predictions (\fig{fig_averaging}d,h,l), even though artifacts can be observed on the edge of the autocorrelations---edges are more sensitive to statistical fluctuations, as they benefit from a weaker spatial averaging effect. 
	\item For $N_{\mathrm{rep}}=40,000$ (acquisition time of 4\,hours), no artifacts due to statistical fluctuations can be observed, and the position of the beads is correctly retrieved (\fig{fig_averaging}c,g,k). 
\end{itemize}
Note that objects with a large number of beads are typically more sensitive to statistical fluctuations, and it can happen that, even for a large number of realizations of the input field, two different beads configurations lead to similar signal autocorrelations that are hard to separate, preventing us to robustly image objects with large numbers of beads using the current reconstruction procedure.


\section{Determination of the number of fluorescent beads}

\begin{figure*}[!b]
	\begin{center}
		\includegraphics[width=0.50\linewidth]{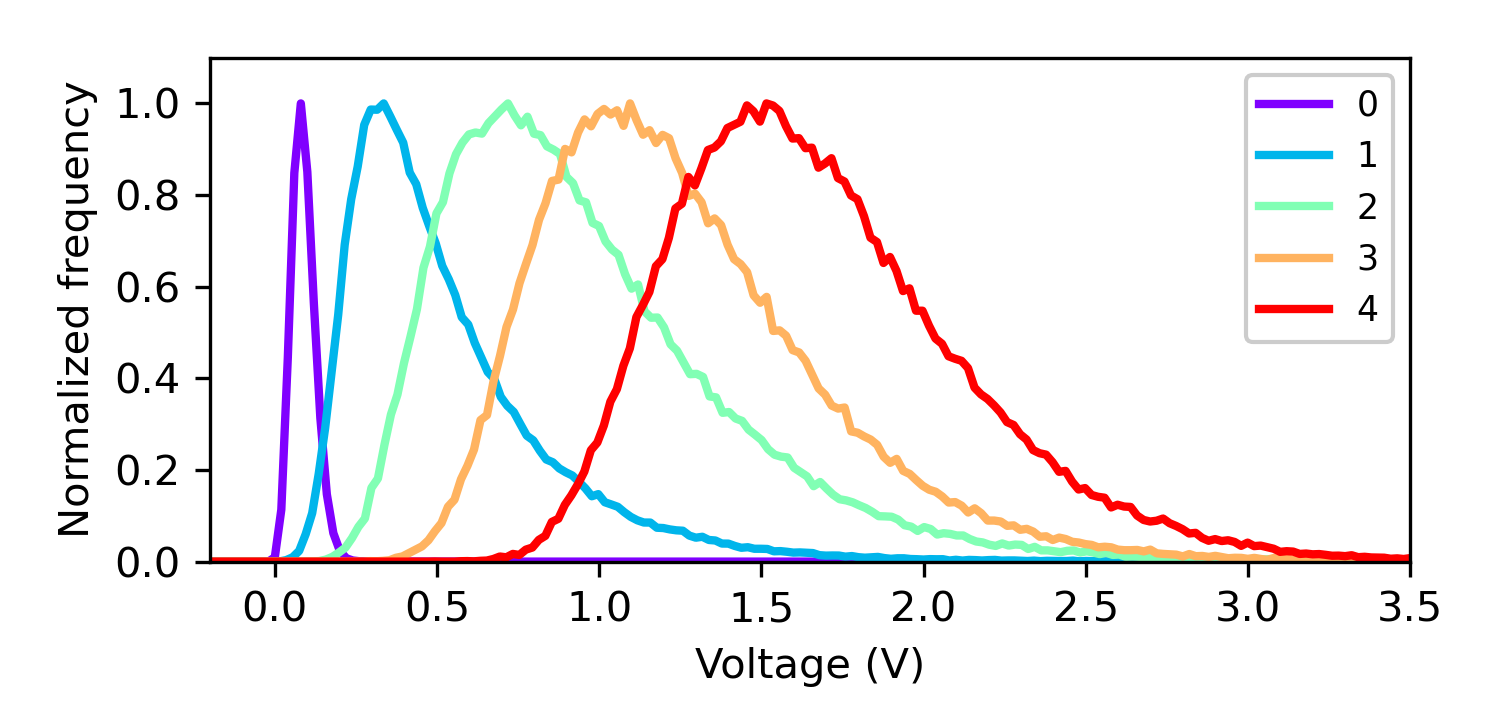}
	\end{center}
	\caption{Statistics of the measured fluorescence signal for different numbers of beads. Voltage distribution associated with the fluorescence signal measured through the $3$\,cm long fiber, for a number of beads ranging from $0$ to $4$. Both the average value and the contrast of these distributions can be used to determine the number of beads. }
	\label{fig_histograms}
\end{figure*}

In principle, the number of fluorescent beads could be determined directly from the measured signal autocorrelation. Nevertheless, a more robust strategy consists in analyzing the statistics of the measured signal. The voltage distribution associated with the fluorescence signal measured through the $3$\,cm long fiber is shown in \fig{fig_histograms}, for the three objects shown in the manuscript (composed of 2, 3 and 4 beads, respectively). For comparison purposes, we also present the distribution obtained in the case of 0 and 1 bead. The average value of these distributions is proportional to the number of beads, with approximately 0.40\,V for each bead. Experimentally, we observed average values of $0.87$\,V for the object composed of 2 beads, $1.17$\,V for the object composed of 3 beads and $1.60$\,V for the object composed of 4 beads, which allowed us to correctly infer the number of beads from these measurements.

Note that another strategy would consist in analyzing the contrast of the measured signals, defined as the ratio between the standard deviation and the average value of the fluorescence signal. In theory, the contrast is equal to $1/\sqrt{n}$ for $n$ point-like fluorescent emitters~\cite{krichevsky_fluorescence_2002_2}. Contrasts measured in our experiment do not reach this theoretical limit, which is expected as the size of the beads ($1.0$\,\textmu m in diameter) is comparable to that of the speckle grain ($0.5 \lambda/ \mathrm{NA}=1.2$\,\textmu m). Nevertheless, observed contrasts do decrease with the number of beads; for the data presented in \fig{fig_histograms}, the contrast is $0.71$ for 1 bead, $0.50$ for 2 beads, $0.37$ for 3 beads and $0.28$ for 4 beads. This indicates that analyzing the contrast is also a possible strategy to determine the number of beads (using e.g. a suitable theoretical model).


\section{Image reconstruction of fluorescent beads from experimental measurements}

Our reconstruction procedure is based on three essential features: the processing of the experimental data, the implementation of the theoretical model, and the optimization algorithm. 

\subsection{Processing of the experimental data}

\begin{figure}[ht]
	\begin{center}
		\includegraphics[width=0.7\linewidth]{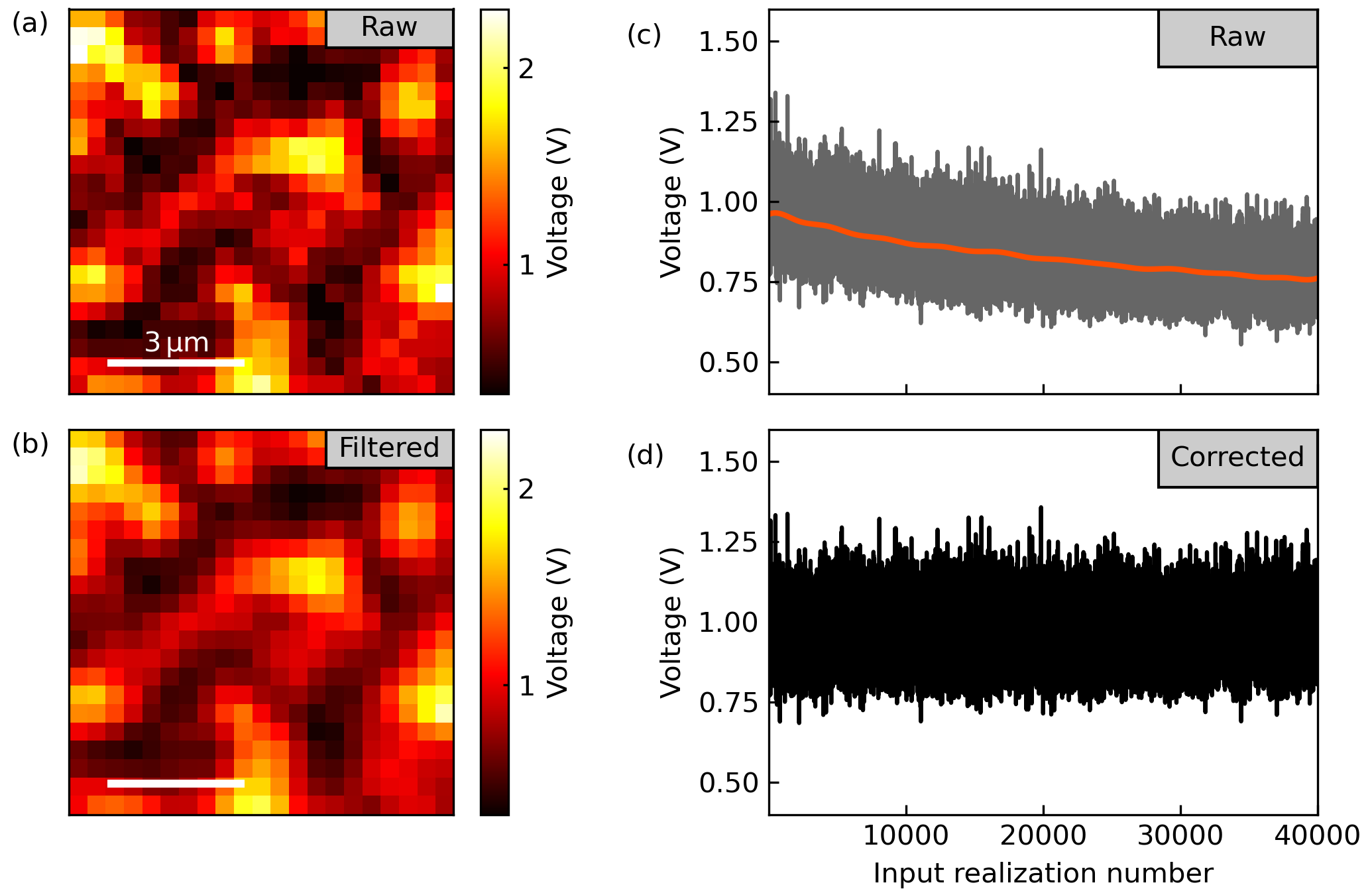}
	\end{center}
	\caption{Processing of measured experimental data. (a) Example of a raw fluorescence signal measured by the photomultiplier as a function of $\Delta \bm{r}_a$, shown here for the object composed of $2$ beads. (b) Same signal as in (a) after being smoothened by a Gaussian spatial filter, that we used to reduce the influence of measurement noise. (c) Average signal measured by the photomultiplier as a function of the number of realizations of the input field (gray curve), along with a model function (orange curve) based on cubic splines. (d) Same signal as in (c) after being compensated for the observed decay. }
	\label{fig_processing}
\end{figure}

The fluorescence signal is experimentally measured for $N_\mathrm{rep}=40,000$ realizations of the input field. For each realization, we sequentially construct a $21$\,$\times$\,$21$ image by translating the input speckle pattern over an area of $8$\,\textmu m\,$\times$\,$8$\,\textmu m and by measuring the resulting intensity with the photomultiplier (\fig{fig_processing}a). This signal is known to be band-limited in the spatial domain due to the finite size of the speckle grain; thus, we apply a Gaussian spatial filter to the measured signal in order to reduce fluctuations arising from measurement noise (\fig{fig_processing}b). Due to the significant acquisition time, we also observed a slow decay of the average signal measured by the photomultiplier, which can be due to laser power fluctuations and to the slow photobleaching of the beads (\fig{fig_processing}c). Therefore, we fit a function based on cubic splines to the temporal dependence of the signal, and we use this function to compensate for this decay (\fig{fig_processing}d). The normalized signal autocorrelation function $C_S(\Delta \bm{r}_a)$ is then directly calculated from its definition [see \eq{eq_cs}], averaging over all possible spatial pairs and all realizations of the input field. Finally, we take advantage of the known symmetry of the signal autocorrelation function and average the calculated signal autocorrelation with its left-right flipped version (or, equivalently, with its up-down flipped version), which further reduces the influence of statistical fluctuations. As a result, we obtain the signal correlation function estimated from experimental measurements, that we denote $C_S^\mathrm{meas}(\Delta \bm{r}_a)$.

\subsection{Implementation of the theoretical model}

In order to theoretically predict the signal autocorrelation from \eq{CS_general}, we must develop a model for the object function $O(\bm{r}_b)$. For this purpose, we model fluorescent beads as high-order Gaussian functions, with the same amplitude for all beads and a full width at half maximum equal to $1$\,\textmu m (i.e. the known diameter of the beads). The object function $O(\bm{r}_b)$ is constructed by summing all contributions from the beads (see Fig.~2b of the manuscript), and its spatial autocorrelation $C_O(\Delta \bm{r}_b)$ is numerically calculated from $O(\bm{r}_b)$ (see Fig.~2c of the manuscript). In addition to $C_O(\Delta \bm{r}_b)$, the theoretical expression of the signal autocorrelation also involves the intensity correlation function $C_I(\Delta \bm{r}_a,\Delta \bm{r}_b)$. While we could use the expression of $C_I(\Delta \bm{r}_a,\Delta \bm{r}_b)$ that was obtained based on an ideal model of square-core fibers [see \eqs{CI_ideal}{Cj}], this would not take into account the limited range of the memory effect that we experimentally observed. For this reason, we estimate $C_I(\Delta \bm{r}_a,\Delta \bm{r}_b)$ from experimental measurements, for both the $3$\,cm long fiber and the $10.5$\,cm long fiber (see Supplementary Section~\ref{sec_CI}). Finally, the signal autocorrelation is calculated from $C_O(\Delta \bm{r}_b)$ and $C_I(\Delta \bm{r}_a,\Delta \bm{r}_b)$ using \eq{CS_general}. As a result, we obtain the predicted signal correlation function (see Fig.~2i of the manuscript), that we denote $C_S^\mathrm{pred}(\Delta \bm{r}_a)$.

\subsection{Optimization algorithm}	

Our reconstruction algorithm is based on the minimization of a loss function that compares theoretical predictions to measured data. As the signal correlation significantly decays with the distance $\Delta \bm{r}_a$, and since the positions of the beads is typically encoded into large values of $\Delta \bm{r}_a$, our loss function is defined using the logarithm of $C_S^\mathrm{meas}(\Delta \bm{r}_a)$ and $C_S^\mathrm{pred}(\Delta \bm{r}_a)$, which efficiently increases the contribution of the most useful parts of the signal correlation function. However, this strategy also tends to increase the contribution of artifacts that are due to statistical fluctuations. Thus, we define a fixed threshold $V_{\mathrm{min}}$ below which values of $C_S^\mathrm{meas}(\Delta \bm{r}_a)$ are considered as artifacts and are not taken into account for the calculation of the loss function. Denoting $\tilde{C}_S^\mathrm{meas}(\Delta \bm{r}_a)$ and $\tilde{C}_S^\mathrm{pred}(\Delta \bm{r}_a)$ the functions ${C}_S^\mathrm{meas}(\Delta \bm{r}_a)$ and ${C}_S^\mathrm{pred}(\Delta \bm{r}_a)$ defined on the restricted domain for which $C_S^\mathrm{meas}(\Delta \bm{r}_a)\geq V_{\mathrm{min}}$, we use the following loss function:
\begin{equation}
	L(\theta) = \left \Vert \log [ \tilde{C}_S^\mathrm{meas}(\Delta \bm{r}_a)] - \log [ \tilde{C}_S^\mathrm{pred}(\Delta \bm{r}_a,\theta) ] \right \Vert^2 + R_\mathrm{fov}(\theta) + R_\mathrm{dmin} (\theta) + R_\mathrm{dmax}(\theta),
	\label{cost}
\end{equation}
where $\Vert \dots \Vert$ denotes the Euclidean distance (i.e. the L2 norm), $\theta=\{x_i,y_i\}_{i=1,\dots,n}$ denotes the positions of the $n$ beads, and $R_\mathrm{fov}(\theta)$, $R_\mathrm{dmin} (\theta)$, and $R_\mathrm{dmax}(\theta)$ denote three regularization terms. The first regularization term $R_\mathrm{fov}(\theta)$ is an exponential potential that penalizes beads located outside the predefined field of view ($10$\,\textmu m\,$\times$\,$10$\,\textmu m). The second regularization term $R_\mathrm{dmin} (\theta)$ is an exponential potential function that penalizes beads that are too close to each other. In this way, we ensure that two beads cannot overlap. The third regularization term $R_\mathrm{dmax} (\theta)$ is an exponential potential function that penalizes beads that are too far apart, based on the area for which $\tilde{C}_S^\mathrm{meas}(\Delta \bm{r}_a)$ is defined. In this way, the distance between two beads is restricted to the area for which measured values of the signal correlation function are significant. 

The cost function defined by \eq{cost} is not convex, and must therefore be minimized using a global optimization strategy. To this end, we implemented an algorithm based on simulated annealing, which is an optimization algorithm inspired by statistical mechanics~\cite{kirkpatrick_optimization_1983_2}. Starting with a random guess for the beads positions, the algorithm typically converges after $200\times p$ iterations, where $p$ is the number of parameters to be estimated (i.e. twice the number of beads since two coordinates must be estimated for each bead). To increase the probability that the global minimum was reached, we repeated this procedure for $20$ different random initial guesses, and we kept the solution that yielded the lowest value of the loss function. In the case of the $3$\,cm long fiber, the number of times that the algorithm converged to this optimal solution was $12/20$ for the object composed of 2 beads, $16/20$ for the object composed of 3 beads, and $3/20$ for the object composed of 4 beads. In the case of the dynamically-perturbed $10.5$\,cm long fiber, the number of times that the algorithm converged to this optimal solution was respectively $2/20$ and $10/20$ for the two objects composed of 4 beads. Overall, this indicates that the inverse problem tends to becomes more difficult to be solved when increasing the number of beads. Note that the algorithm frequently reconstructs flipped versions of the objects, which is expected due to the known ambiguity in the inverse problem (see Supplementary Section~\ref{sec_ambiguities}). In the manuscript, reconstructed images were presented by selecting the version that best corresponds to the ground truth among the 4 possibilities. 


\section{Speckle decorrelation induced by perturbing the fiber}

\label{sec_decorrelation}

\begin{figure}[!b]
	\begin{center}
		\includegraphics[width=0.78\linewidth]{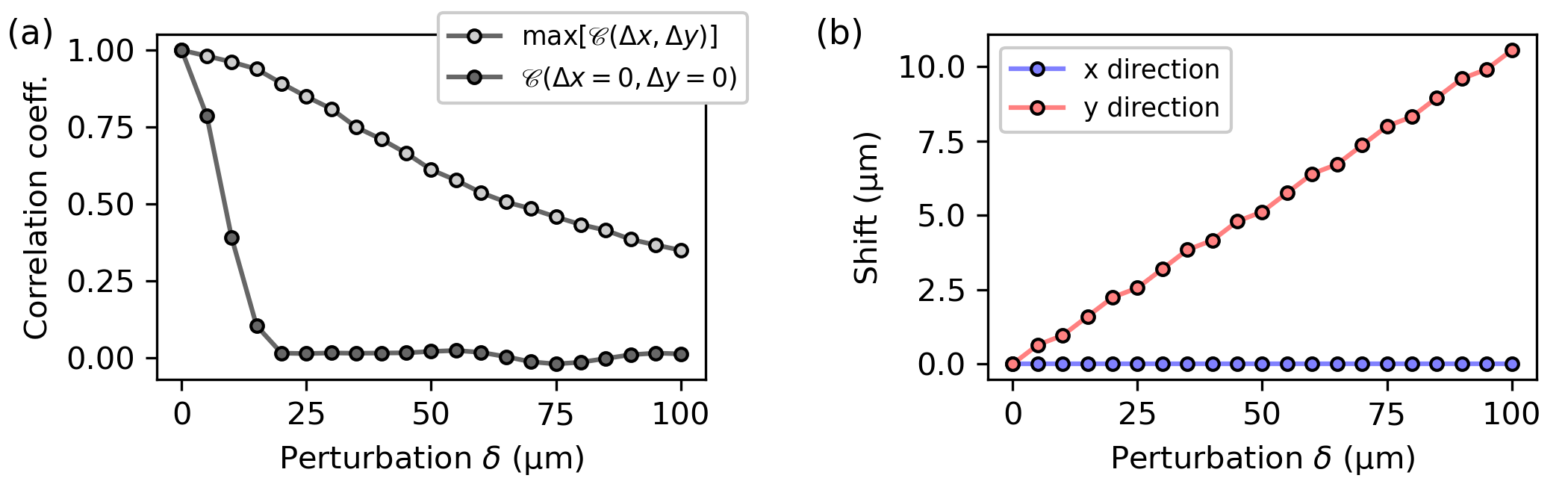}
	\end{center}
	\caption{(a) Correlation coefficient as a function of the applied perturbation, taking as a reference the speckle pattern measured for $\delta=0$\,\textmu m. The correlation coefficient evaluated at zero shift $\mathscr{C}(\Delta x=0,\Delta y=0)$ includes the effect of speckle decorrelation and of the transverse shift of the fiber output, while the maximum of the correlation coefficient $\mathrm{max}[\mathscr{C}(\Delta x,\Delta y)]$ includes only the effect of speckle decorrelation. (b) Transverse shift of the fiber output as a function of the applied perturbation, calculated from the position of the maximum value of $\mathscr{C}(\Delta x,\Delta y)$.}
	\label{fig_decorrelation}
\end{figure}

In order to quantitatively assess the influence of the perturbation induced by the rod upon the transmission matrix of the imaging system, we generate a random speckle pattern at the fiber input and we measure output speckle patterns for different displacements $\delta$ of the rod. We then calculate the spatial cross-correlation $\mathscr{C}(\Delta x, \Delta y)$ of the measured patterns for these different values of $\delta$, taking as a reference the pattern measured for $\delta=0$\,\textmu m. The value of $\mathscr{C}(\Delta x=0, \Delta y=0)$ decreases from one to zero for a displacement of the rod of approximately $20$\,\textmu m (\fig{fig_decorrelation}a, dark points), evidencing that the transmission matrix of the imaging system is completely modified by a perturbation $\delta \geq 20$\,\textmu m. 

The observed perturbation of the transmission matrix is due not only to a decorrelation of the speckle patterns but also to a transverse shift of the fiber. In order to disentangle the influence of these two effects, we calculate the maximum value of the function $\mathscr{C}(\Delta x, \Delta y)$ for each value of the perturbation $\delta$ (\fig{fig_decorrelation}a, light points). We observe that the value of $\mathrm{max}[\mathscr{C}(\Delta x, \Delta y)]$ also decreases with $\delta$, reaching a value of $0.35$ for $\delta=100$\,\textmu m. This indicates that the modification of the transmission matrix of the imaging system is not entirely due to the decorrelation of the speckle patterns, but that it is also partly due to a transverse shift of the fiber. This is confirmed by studying the shift of the fiber output as a function of $\delta$ (\fig{fig_decorrelation}b), which occurs in the direction of the displacement of the rod inducing the perturbation (the $y$ direction) and reaches a value of $10.6$\,\textmu m for $\delta=100$\,\textmu m. 

We emphasize that, since both the decorrelation of the speckle patterns and the transverse shift of the fiber modify the transmission matrix of the imaging system, the relevant metric to quantify the influence of the perturbation is here $\mathscr{C}(\Delta x=0, \Delta y=0)$, which is represented by the dark points in \fig{fig_decorrelation}a and which is also shown in Fig.~4e of the manuscript.


\section{Image reconstruction through a 10.5 cm long fiber with and without dynamic perturbations}

\begin{figure}[ht]
	\begin{center}
		\includegraphics[width=0.78\linewidth]{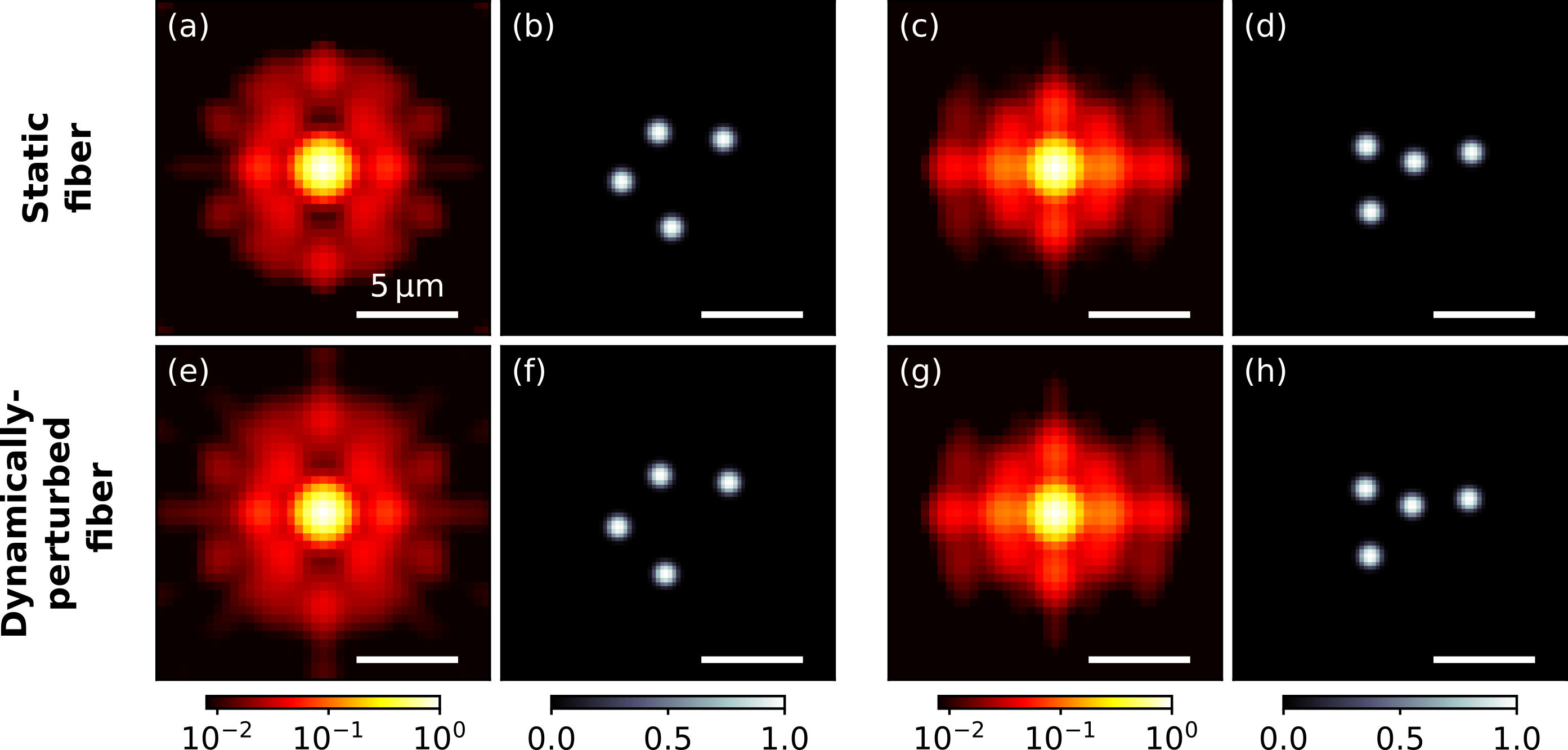}
	\end{center}
	\caption{(a) Signal autocorrelation retrieved from experimental measurements (averaged over $N_{\mathrm{rep}}=40,000$ speckle illuminations), for the object presented in Fig.~ 4g of the manuscript. For these measurements, the 10.5 cm long fiber was maintained in a static position. (b) Reconstructed images of the beads. (c-d) Analogous to (a-b) for the object presented in Fig.~4h of the manuscript. (e-h) Signal autocorrelations and reconstructed images of the beads from experimental measurements performed while dynamically perturbing the fiber. Note that these results are those presented from Fig.~4k to Fig.~4n in the manuscript.}
	\label{fig_dynamic}
\end{figure}

To further demonstrate that the dynamical aspect of the applied perturbation does not significantly influence the efficiency of the method, we present here complementary experimental results obtained from the objects that we presented in Fig.~~4 of the manuscript. However, instead of dynamically-perturbing the fiber, we maintained the fiber in a static position (\fig{fig_dynamic}, top row). Both the measured signal autocorrelations and the reconstructed images are very similar to the results obtained using a dynamically-perturbed fiber (\fig{fig_dynamic}, bottom row). This confirms that, while some stability is required when scanning a given input speckle pattern ($220$\,ms in our experiments), the fiber can be perturbed between different random realizations of the input speckle pattern without affecting the efficiency of the method.

\section{Image reconstruction of handwritten digits from numerical simulations}

\label{sec_ANN}

\begin{figure}[ht]
	\begin{center}
		\includegraphics[width=\linewidth]{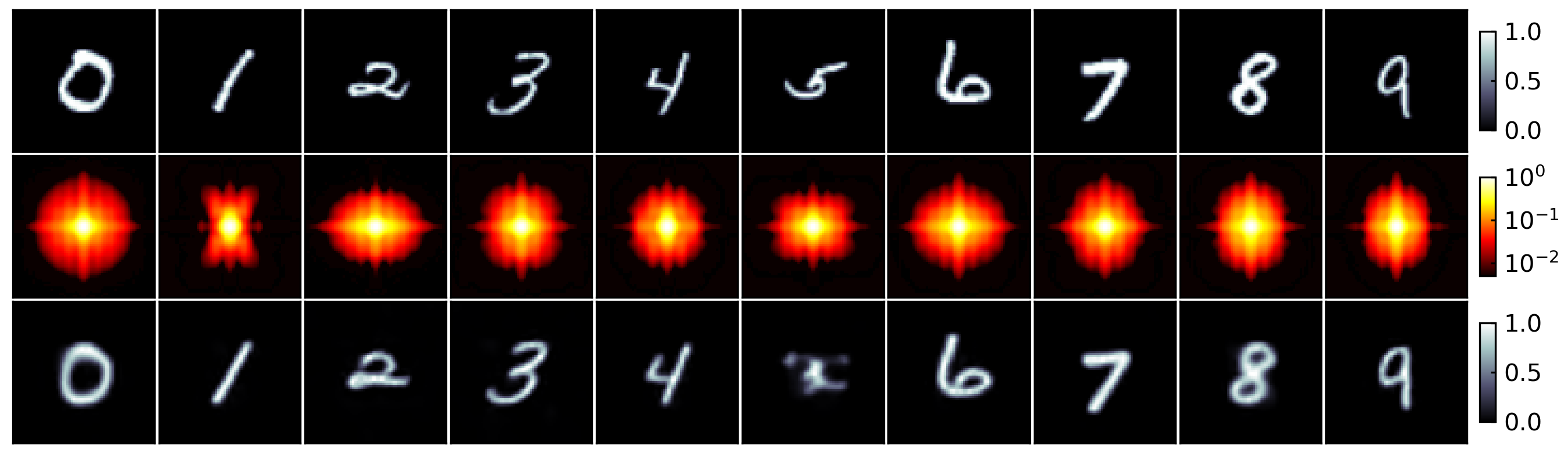}
	\end{center}
	\caption{Reconstructed images of handwritten digits in numerical simulations (see also Fig.~4 of the manuscript). First row: Grayscale images from the MNIST database of handwritten digits, that are used as objects in our numerical simulations. Second row: Signal autocorrelations of these objects calculated using \eq{CS_general}. Third row:~Images reconstructed by an artificial neural network, demonstrating that the inverse problem can be successfully solved even in the case of complicated objects. In one of these examples (the one associated with the digit 5), the image was not properly reconstructed, illustrating the fact that the procedure is not error-free.}
	\label{fig_ANN}
\end{figure}

To complement the results presented in Fig.~4 of the manuscript, we show in \fig{fig_ANN} additional simulation results. In this figure, the top row is composed of objects extracted from the test set, the middle row is composed of the associated signal autocorrelations, and the bottom row is composed of the predicted images reconstructed by the artificial neural network. In one of these examples (the one associated with the digit 5), the image was not properly reconstructed, illustrating the fact that the procedure is not error-free. However, over the whole test set, the average structural similarity is $0.89$, which indicates that the predictions generally strongly resemble the ground truths. In addition, we trained a classifier using signal autocorrelations as inputs, which yields a success rate of $91$\%. Overall, these results demonstrate the strong potential of artificial neural networks to successfully solve the inverse problem and reconstruct images based on measured signal autocorrelations.

\end{document}